\documentclass[doublespacing,onecolumn,extra]{gji}

\usepackage{timet}
\usepackage{amsfonts}
\usepackage{mathrsfs}
\usepackage{graphicx}
\usepackage{bm}
\usepackage{caption}
\usepackage{subcaption}
\usepackage{ amssymb }
\usepackage{tikz-cd}
\usepackage{csquotes}

\title[Adjoint GIA]
{Application of first- and second-order adjoint methods to glacial isostatic adjustment incorporating rotational feedbacks}
\author[Z. Yu \emph{et al.}]
{
   Ziheng Yu$^{1}$, David Al-Attar$^{1}$, Frank Syvret$^{1}$, and Andrew J. Lloyd$^{2}$ \\
  $^{1}$University of Cambridge, Bullard Laboratories, Madingley Road, Cambridge CB3 OEZ, UK. Email: zy296@cam.ac.uk   \\
  $^{2}$Lamont Doherty Earth Observatory, Columbia University, Palisades, NY 10964, USA.
}

\date{Received ?; in original form ?}
\pagerange{\pageref{firstpage}--\pageref{lastpage}}
\volume{142}
\pubyear{2020}
\def\LaTeX{L\kern-.36em\raise.3ex\hbox{{\small A}}\kern-.15em
  T\kern-.1667em\lower.7ex\hbox{E}\kern-.125emX}

\newcommand{\unvec}{\hat{\mathbf{n}}}

\newcommand{\diff}[1]{\left[#1\right]^{+}_{-}}

\newcommand{\dd}{\,\mathrm{d}}
\newcommand{\ddns}{\mathrm{d}}

\newcommand{\vertiii}[1]{{\left\vert\kern-0.25ex\left\vert\kern-0.25ex\left\vert
      #1 \right\vert\kern-0.25ex\right\vert\kern-0.25ex\right\vert}}

\begin{document}

\label{firstpage}

\maketitle

\begin{summary}
  This paper revisits and extends  the adjoint theory for glacial isostatic adjustment (GIA) of
  \cite{crawford2018}. Rotational feedbacks are now incorporated, and the application 
  of the second-order adjoint method is described for the first time. The first-order adjoint 
  method provides an efficient means for computing sensitivity kernels for a chosen 
  objective functional, while the second-order adjoint method provides second-derivative information
  in the form of Hessian kernels. These latter kernels are required by 
  efficient Newton-type optimisation schemes and  within
     methods for quantifying uncertainty  for non-linear
  inverse problems.  Most importantly, the entire theory has been reformulated so as to simplify its
  implementation by others within the GIA community.
  In particular, the  rate-formulation for the GIA forward problem introduced by  \cite{crawford2018} 
  has  been replaced with the conventional equations  for modelling GIA in laterally heterogeneous
  earth models. The implementation of the first- and second-order adjoint problems should be 
  relatively easy within both  existing and new GIA codes, with only the inclusions of more general force terms 
  being required.   
\end{summary}


\section{Introduction}

In broad terms, the GIA inverse problem uses palaeo sea level records and related observations (e.g., present-day geodetic velocities and 
  gravity anomalies)
to quantitatively reconstruct ice sheet history back into the last glacial whilst
simultaneously estimating mantle viscosity variations \citep[e.g.][]{peltierinv,peltiervm2,lambeck,peltiervm5,arguspeltier,nakadaokuno,roy2018relative,li2020uncertainties}.
Work on this  problem has  relied largely on  guided forward modelling, but
such an approach is feasible only if mantle viscosity is assumed to vary  as function of depth alone. The importance of 
lateral viscosity variations for GIA has now been clearly established \citep[e.g.][]{latychev2005gia,austermann2013barbados,van2015effect,lau2018inferences,pan2022influence}, though their inclusion within GIA modelling raises 
the computational cost dramatically \citep[e.g.][]{zhong,latychev,steffen2006three,wahr2013computations}.
Experience gained within other fields concerned with large-scale inverse problems \citep[e.g.][]{lions,biegler2003large,tapeliutromp,troltzsch,biros2005parallelI, biros2005parallelII,
 wunsch2006discrete,fichtner2006adjoint,metivier2013full,bozdaug2016global} suggests that 
the application of adjoint methods to GIA will be essential in  making  progress with the inverse problem  while accounting for lateral viscosity variations.
 Adjoint methods provide an optimally efficient means for determining the gradients   within iterative optimisation schemes \citep[e.g.][]{nocedal2006numerical},
and they also  form an essential component within the  application of Bayesian methods to large-scale non-linear inverse problems \citep[e.g.][]{stuart2010inverse,bui2013computational,petra2014computational,
papadimitriou2015bayesian}. The development of surrogate models for GIA in laterally heterogeneous earth models \citep[e.g.][]{love2023fast} may change this outlook, but their viability
 is yet to be  firmly established.

 Theoretical  aspects of the adjoint method in the context of GIA have been discussed in a number of papers 
 \citep[][]{alattartromp,martinec2015forward,crawford2018} while 
 others have focused on initial applications \citep[][]{kim2022ice,lloyd2023}.
 The aim of this work  is to revisit and extend the approach of
 \cite{crawford2018} which  is the most complete version of 
 the adjoint theory to date.  The first  point we address is the incorporation
 of rotational feedbacks into the adjoint problem.   The next  
 is the development of  second-order adjoint theory. First-order adjoint theory allows for efficient calculation 
 of the  derivatives of a chosen scalar-valued 
 objective functional, with  these derivatives being required
 within gradient-based optimisation schemes. Similarly, the second-order 
 adjoint theory allows for the action of the Hessian operator (i.e., the 
 second derivative of the objective functional) on a given model perturbation to 
 be efficiently determined. Such calculations are required in the application 
 of Newton-type optimisation schemes that offer superior convergence properties
 \citep[e.g.][]{nocedal2006numerical,metivier2013full}. Hessian calculations 
 are also necessary within a range of  methods for quantifying uncertainties within 
 large-scale and non-linear inverse problems \citep[e.g.][]{wang1992second,fichtner2011hessian,
 bui2013computational,petra2014computational,papadimitriou2015bayesian}.

 Beyond the extensions just mentioned, this paper  presents a complete reformulation 
 of the adjoint theory of \cite{crawford2018} with the aim of making these methods more 
 readily understandable and applicable within the GIA community. \cite{crawford2018}
 built on the rate formulation for viscoelastic loading developed in \cite{alattartromp}.
 An appealing feature of the rate-formulation is that the GIA forward problem can be written as a 
 coupled system of non-linear evolution equations without need for the usual iterative
 solution of the sea level equation. The rate-formulation also offers certain advantages
 in terms of the adjoint problem. In particular, within this approach
 the rate of change of the ice thickness occurs naturally as a model parameter, and 
 hence  singular behaviour of the sensitivity kernel for ice thickness is 
 avoided. Within the present work, by contrast,  singularities of the ice kernel
 must be addressed directly. Nevertheless, the rate-formulation  is not used within other
 GIA codes that account for laterally varying viscosity \citep[e.g.][]{zhong,latychev,steffen2006three}
 and its adoption within them would require non-trivial modifications.
  
 A further point is that the numerical implementation of the rate-formulation  depends on 
 explicit time-stepping schemes that lack uniform stability. The largest time-step
 is set by a constant fraction of the smallest relaxation time within the earth model. For calculations
 in realistic laterally heterogeneous models, this time-step can be prohibitively
 small for simulations spanning the deglacial period. While there do exist
 stable time-stepping schemes suitable for modelling quasi-static
 viscoelastic deformation \citep[e.g.][]{simohughes,bailey2006large}, their implementation
 within the context of the rate-formulation is challenging and may not be possible. 
 By moving towards a conventional formulation of the GIA forward problem, these numerical difficulties can be circumvented, and hence the application 
 of adjoint methods within the GIA inverse problem made significantly more efficient.

\section{Summary of the forward GIA problem}

We begin by summarising in suitable form the equations of motion for GIA. In
doing this, we assume a Maxwell rheology and account for shoreline migration
and rotational feedbacks. This discussion builds directly on
\cite{alattartromp}, \cite{ crawford2018}, and \cite{al2024reciprocity}, with
these works being in turn based on the earlier literature
\citep[e.g.][]{peltier74,dahlen74,farrellclark,peltierinv,mitrovica1991postglacial,dt,
  milnemitrovica98,trompmitrovica,mitrovicamilne,kendall}. For simplicity, we
neglect the existence of a fluid outer core within the main text. The necessary 
details are provided within Appendix A where it is shown that
the inclusion of fluid regions has no effect on the form 
of our main results.  

\subsection{Static loading on an elastic planet}

We consider a non-rotating and isotropic elastic earth model that is initially
in hydrostatic equilibrium. The earth model does not include an ocean or ice
sheets, with these features later introduced through appropriate surface loads.
Let $M$ denote the volume of the earth model at equilibrium, and $\partial M$ its surface which has outward unit normal $\unvec$. The hydrostatic
equilibrium condition requires that
\begin{equation}
  \nabla p + \rho \nabla \Phi = 0,
\end{equation}
where $p$, $\rho$, and $\Phi$ are, respectively, the pressure, density, and gravitational potential.

Suppose that the equilibrium state is disturbed by the application of a surface
load, $\sigma$. The resulting deformation can be described by a displacement
vector $\mathbf{u}$, along with an Eulerian perturbation, $\phi$, to the
gravitational potential. The linearised equations of motion can be written
concisely in weak form as
\begin{equation}
  \label{eq:EQM1}
  \mathcal{A}(\mathbf{u},\phi \,| \, \mathbf{u}', \phi') + \int_{\partial M} (\mathbf{u}'\cdot \nabla \Phi + \phi') \,\sigma \dd S = 0,
\end{equation}
which is required to hold for arbitrary  test functions  $\mathbf{u}'$ and $\phi'$ \citep{alattartromp}. Here we
have the bilinear form 
\begin{eqnarray}
  \mathcal{A}(\mathbf{u},\phi \,| \, \mathbf{u}', \phi') &=& \int_{M} \kappa\, \nabla \cdot \mathbf{u} \, \nabla \cdot \mathbf{u}' \dd^{3}\mathbf{x}
  + \int_{M}2 \mu\, \mathbf{d}:\mathbf{d}' \dd^{3} \mathbf{x} + \frac{1}{2}\int_{M}\rho \left[
    \nabla(\mathbf{u}\cdot \nabla \Phi)\cdot \mathbf{u}' + \nabla(\mathbf{u}'\cdot \nabla \Phi)\cdot \mathbf{u}
    \right] \dd^{3} \mathbf{x} \nonumber \\
  && -\frac{1}{2}\int_{M}\rho \left(
  \nabla \cdot \mathbf{u} \, \nabla \Phi \cdot \mathbf{u}' + \nabla \cdot \mathbf{u}' \, \nabla \Phi \cdot \mathbf{u}
  \right) \dd^{3}\mathbf{x} + \int_{M} \rho \left(
  \nabla \phi  \cdot \mathbf{u}' + \nabla \phi'  \cdot \mathbf{u}
  \right) \dd^{3} \mathbf{x} \nonumber \\ && + \frac{1}{4\pi G}\int_{\mathbb{R}^{3}} \nabla \phi \cdot \nabla \phi' \dd^{3}\mathbf{x},
\end{eqnarray}
associated with elastic and gravitational restoring forces.
Within this expression, $G$ is the gravitational constant, and  $\kappa$ and $\mu$ are the bulk and shear
modulii. The linearised strain tensor, $\mathbf{e}$  is defined by
\begin{equation}
  \mathbf{e} = \frac{1}{2}\left[\nabla \mathbf{u} + (\nabla \mathbf{u})^{T}\right],
\end{equation}
while its deviatoric part is
\begin{equation}
  \mathbf{d} = \mathbf{e} - \frac{1}{3}\mathrm{tr}(\mathbf{e}) \mathbf{1},
\end{equation}
with $\mathrm{tr(\cdot)}$ denoting the trace of a matrix and  $\mathbf{1}$ the identity matrix. The term $\mathbf{d'}$ is
defined in an identical manner with respect to the test function $\mathbf{u}'$.
We note the symmetry
\begin{equation}
  \label{eq:Aadjoint}
  \mathcal{A}(\mathbf{u},\phi \,| \, \mathbf{u}', \phi') = \mathcal{A}(\mathbf{u}',\phi' \,|
  \, \mathbf{u}, \phi),
\end{equation}
that will be used repeatedly later on. As a final condition, we require that $\phi$ tends to zero at 
infinity, this serving to fix the arbitrary constant that can always be added to the potential.

\subsection{Rigid body motions}

A linearised rigid body motion takes the form
\begin{equation}
  \mathbf{u} = \mathbf{a} + \mathbf{b} \times \mathbf{x},
\end{equation}
with $\mathbf{a}$ and $\mathbf{b}$ constant vectors that describe, respectively,
degrees of freedom associated with translation and rotation. The associated
gravitational potential perturbation at a fixed spatial point is
\begin{equation}
  \phi = - (\mathbf{a} + \mathbf{b} \times \mathbf{x})\cdot \nabla \Phi.
\end{equation}
For such fields it can be verified directly that
\begin{equation}
  \label{eq:Akernel}
  \mathcal{A}(\mathbf{u},\phi \,| \, \mathbf{u}', \phi') = 0,
\end{equation}
for any test functions $(\mathbf{u}', \phi')$.
An immediate consequence is that the solution of the static elastic loading
problem is defined only up to an arbitrary linearised rigid-body motion. This
means that if $(\mathbf{u},\phi)$ solves the loading problem for a given
$\sigma$, then so do the fields
\begin{equation}
  \mathbf{u} + \mathbf{a} + \mathbf{b} \times \mathbf{x}, \quad
  \phi - (\mathbf{a} + \mathbf{b} \times \mathbf{x})\cdot \nabla \Phi,
\end{equation}
for any $\mathbf{a}$ and $\mathbf{b}$. Assuming that the elastic earth model is gravitationally stable, 
it can be shown that the six-dimensional subspace of linearised rigid body motions comprises
the only fields for which eq.(\ref{eq:Akernel}) holds \citep[e.g.][Sections 4.1.5 and 4.1.6]{dt}.

Suppose  that within eq.(\ref{eq:EQM1}) we take as test functions
\begin{equation}
  \mathbf{u}' = \mathbf{a} + \mathbf{b} \times \mathbf{x}, \quad   \phi' = - (\mathbf{a} + \mathbf{b} \times \mathbf{x})\cdot \nabla \Phi.
\end{equation}
Using eq.(\ref{eq:Aadjoint}) and (\ref{eq:Akernel}),  the first term within eq.(\ref{eq:EQM1}) vanishes,
and hence we are left with 
\begin{equation}
\int_{\partial M} (\mathbf{u}'\cdot \nabla \Phi + \phi') \,\sigma \dd S = 0,
\end{equation}
as a necessary condition on the load in order for solutions of the problem to exist. 
In fact, it can readily be seen that $\mathbf{u}'\cdot \nabla \Phi + \phi' = 0$ for our chosen test functions,
and hence the  condition on the load is automatically met.  
It follows from  the Fredholm alternative for elliptic partial differential equations
 that the above condition on the load is also sufficient for  solutions of eq.(\ref{eq:EQM1}) to exist \citep[e.g.][Chapter 6]{marsden1994mathematical}.
We conclude that the static loading problem has solutions for any given load, with these solutions defined uniquely up to the 
addition of an arbitrary linearised rigid motion.

\subsection{Rotational feedbacks}

Surface loading of the Earth is associated with changes to its moment of
inertia and hence to its angular velocity to conserve angular momentum. These
rotational changes in turn generate centrifugal forces that modify the
deformation, and thus feedbacks exist between load-induced deformation and
rotational variations \citep[e.g.][]{sabadini1981pleistocene,
  wu1984pleistocene,milnemitrovica98,martinec2005time,mitrovica2005rotational,
  mitrovica2011ice}. Within GIA studies, it has been usual to account for
rotational feedbacks in an approximate manner as summarised below. A
notable feature of this approach is that it allows for calculations to be
performed in spherically symmetric earth models which, until relatively
recently, was all that was feasible. 

Within the traditional approach to rotational feedbacks, static loading of a non-rotating model
is still considered, but an additional force term is included to represent a
centrifugal potential perturbation. Following the discussion in
\cite{al2024reciprocity}, the static elastic loading problem in eq.(\ref{eq:EQM1}) is generalised to read
\begin{equation}
  \label{eq:EQM2}
  \mathcal{A}(\mathbf{u},\phi \,| \, \mathbf{u}', \phi') + \int_{\partial M} (\mathbf{u}'\cdot \nabla \Phi + \phi') \,\sigma \dd S +\int_{M}\rho \,
  \mathbf{u}' \cdot \nabla \psi \dd^{3}\mathbf{x}= 0,
\end{equation}
with $(\mathbf{u}',\phi')$ again  test functions, and where
the perturbed centrifugal potential, $\psi$,  is related to the perturbed angular velocity, $\bm{\omega}$, through
\begin{equation}
  \psi = -(\bm{\Omega}\times \mathbf{x}) \cdot (\bm{\omega}\times \mathbf{x}),
\end{equation}
with $\bm{\Omega}$ the equilibrium angular velocity. For given values of the load, $\sigma$, and
centrifugal potential perturbation, $\psi$, the linearised equations of motion
can be solved for the displacement vector $\mathbf{u}$ and the gravitational
potential perturbation, $\phi$. Using these results along with the direct
contribution of the surface load, the perturbation to the model's inertia
tensor can be determined. From the inertia tensor perturbation the resultant
change in the angular velocity can be found via conservation of angular
momentum. The latter calculation depends on knowledge of the equilibrium
inertia tensor, though we note that this tensor cannot, in general, be taken
directly from the earth model in which the elastic calculations have been
performed \citep[e.g.][]{mitrovica2005rotational}. As shown in \cite{al2024reciprocity}, the resulting 
relationship between the displacement, load, and perturbed angular
velocity can be concisely written
\begin{equation}
  \label{eq:Inertia}
  \bm{\omega}' \cdot \mathbf{D}
\cdot \bm{\omega} + \int_{M} \rho\,\mathbf{u} \cdot \nabla\psi' \dd^{3}\mathbf{x} + \int_{\partial M} \sigma \,\psi' \dd S = 0,
\end{equation}
where $\bm{\omega}'$ is an arbitrary vector, $\psi' =
  -(\bm{\Omega}\times \mathbf{x}) \cdot (\bm{\omega}'\times \mathbf{x})$, and $\mathbf{D}$
the symmetric and invertible matrix
\begin{equation}
  \mathbf{D} = \left(
  \begin{array}{ccc}
      C_{3} -C_{1} & 0           & 0      \\
      0            & C_{3}-C_{2} & 0      \\
      0            & 0           & -C_{3}
    \end{array}
  \right),
\end{equation}
with $C_{1}\le C_{2} < C_{3}$ the principal moments of inertia assumed.
Equations (\ref{eq:EQM2}) and (\ref{eq:Inertia}) can be combined into a single
weak form
\begin{equation}
  \label{eq:EQM3}
  \mathcal{A}(\mathbf{u},\phi \,| \, \mathbf{u}', \phi') + \int_{\partial M} (\mathbf{u}'\cdot \nabla \Phi + \phi' + \psi') \,\sigma \dd S
  +\int_{M}\rho \left(\mathbf{u}' \cdot \nabla \psi + \mathbf{u} \cdot \nabla
  \psi' \right) \dd^{3}\mathbf{x} + \bm{\omega}' \cdot \mathbf{D} \cdot
  \bm{\omega} = 0,
\end{equation}
which is to   hold for arbitrary  test functions  $\mathbf{u}'$, $\phi'$, and $\bm{\omega}'$.
While these equations could  be solved for $\mathbf{u}$, $\phi$, and $\bm{\omega}$ simultaneously,  
it is typical for an iterative approach to be applied with an initial guess for $\bm{\omega}$
successively refined.

\subsection{Rigid body motions (again)}

We need to reassess the issue of uniqueness and existence within the loading
problem now that rotational feedbacks have been included. As with the earlier
static loading problem, for given $(\sigma, \psi)$ we see from
eq.(\ref{eq:EQM2}) that solutions can only be defined up to a linearised rigid
body motion. If we take
\begin{equation}
  \mathbf{u}' = \mathbf{a} + \mathbf{b} \times \mathbf{x}, \quad   \phi' = - (\mathbf{a} + \mathbf{b} \times \mathbf{x})\cdot \nabla \Phi,
\end{equation}
within eq.(\ref{eq:EQM2})  we arrive at the equality
\begin{equation}
  \label{eq:rotextmp}
  \int_{M} (\mathbf{a} + \mathbf{b} \times \mathbf{x})\cdot \rho\,\nabla \psi \dd^{3}\mathbf{x} = 0,
\end{equation}
as a necessary and sufficient condition for solutions to exist. Assuming for the
moment that this condition  is met, we can find a solution $(\mathbf{u},\phi)$ of the
static elastic equations determined up to a linearised rigid body motion.  The matrix $\mathbf{D}$
in eq.(\ref{eq:Inertia}) is invertible, and hence this
equation has a unique solution, $\bm{\omega}$, for   given $(\mathbf{u},\sigma)$. Because, however,
$\mathbf{u}$ is only determined by eq.(\ref{eq:EQM2}) up to a linearised rigid
body motion, we need to ask how this non-uniqueness manifests in $\bm{\omega}$.
If, within, eq.(\ref{eq:Inertia}) we replace the displacement vector
by $\mathbf{u} +\mathbf{a} + \mathbf{b} \times \mathbf{x} $, it can be seen that
the solution $\bm{\omega}$ is unchanged if and only if
\begin{equation}
  \label{eq:rotex}
  \int_{M} (\mathbf{a} + \mathbf{b} \times \mathbf{x})\cdot \rho\,\nabla \psi' \dd^{3}\mathbf{x} = 0,
\end{equation}
where we recall that $\psi'$ is the centrifugal potential perturbation associated with
an arbitrary angular velocity $\bm{\omega}'$; clearly this requirement is equivalent to the
existence condition in eq.(\ref{eq:rotextmp}).
Summarising the above discussion, we have shown that solutions of the elastostatic
loading problem with rotational feedbacks exist so long as
eq.(\ref{eq:rotex}) holds for arbitrary $\bm{\omega}'$. When this condition is met, the solutions
are  defined uniquely modulo the transformation
\begin{equation}
  \mathbf{u} \mapsto \mathbf{u} +  \mathbf{a} + \mathbf{b} \times \mathbf{x}, \quad
  \phi \mapsto \phi  - (\mathbf{a} + \mathbf{b} \times \mathbf{x})\cdot \nabla \Phi,
  \quad \bm{\omega} \mapsto \bm{\omega},
\end{equation}
for arbitrary constant vectors $\mathbf{a}$ and $\mathbf{b}$.

It remains to determine whether eq.(\ref{eq:rotex}) holds in practice. To do this,  consider
eq.(\ref{eq:Inertia}) in the case that $\sigma =0$. We then have
\begin{equation}
  \bm{\omega}' \cdot \mathbf{D} \cdot \bm{\omega} + \int_{M} \rho\,\mathbf{u} \cdot \nabla\psi' \dd^{3}\mathbf{x}  = 0,
\end{equation}
which gives the perturbed angular velocity, $\bm{\omega}$, associated with the
displacement vector, $\mathbf{u}$. If we take $\mathbf{u} = \mathbf{a} + \mathbf{b} \times \mathbf{x} $
within this equation and assume that eq.(\ref{eq:rotex}) holds, then we see that $\bm{\omega} = \mathbf{0}$.
It follows that eq.(\ref{eq:rotex}) is equivalent to the invariance of the inertia tensor
for the equilibrium earth model under arbitrary linearised rigid body motions, and hence  this tensor must be
isotropic. Note that this is a statement about the earth model assumed within the elastostatic
calculations, and not about the equilibrium inertia tensor used when
calculating perturbations to the angular velocity via conservation of angular momentum. 

We have shown that the standard theory of rotational feedbacks implicitly assumes that model in which the
elastic calculations are performed has an isotropic moment of inertia tensor.
 To our knowledge, this is an original observation. This rotational theory was  developed 
within the context of spherically symmetric earth models for which the requirement is met, 
but it remains in use within more recent calculations in
laterally heterogeneous earth models and here problems could emerge. In
practice, however, such calculations are focused on the effect of large lateral
viscosity variations within the mantle, with smaller lateral variations in the
model's shape or density typically being ignored. More sophisticated theories
for rotational feedbacks have been developed based upon perturbations to a
rotating equilibrium state, and  there then exist no implicit
constraints on the structure of the earth model. Work in this direction includes
\cite{martinec2005time}, \cite{martinec2014rotational} along with the recent
study by \cite{maitra2024elastodynamics} which includes a complete non-linear
theory for quasi-static loading within a variably rotating earth model. Within
this paper, we nonetheless retain the traditional rotational theory because it
is the most widely used and because it remains appropriate in most cases of practical
interest.

\subsection{Viscoelastic relaxation}

The equations stated so far apply to the static loading of an elastic earth
model, but they can be readily extended to the case of quasi-static
viscoelasticity. Within GIA studies, a Maxwell solid rheology is most commonly
assumed, and we focus on this case within this paper.
The incorporation of more complex linear or non-linear rheolgies within 
both the forward and adjoint theory is not difficult; see \cite{crawford2017}
for results along these lines in the context of post-seismic deformation. 
As discussed by \cite{alattartromp}, the linearised stress tensor
for a Maxwell solid is
\begin{equation}
  \label{eq:Maxwell}
  \mathbf{T} = \kappa \,\nabla \cdot \mathbf{u}\, \mathbf{1} + 2\mu \,(\mathbf{d} - \mathbf{m}),
\end{equation}
where $\mathbf{m}$ is an internal variable that satisfies the following differential equation
\begin{equation}
  \dot{\mathbf{m}} + \frac{1}{\tau}(\mathbf{m}-\mathbf{d}) = \mathbf{0},
\end{equation}
with $\tau$ the relaxation time. Such an internal variable
approach to viscoelasticity is widely used within the engineering literature
\citep[e.g.][]{simohughes,nsm} and it is equivalent to other time-domain
schemes that have been applied within the context of GIA
\citep[e.g.][]{hanyk95,zhong,latychev,bailey2006large}.
Generalising eq.(\ref{eq:EQM3}) to account for Maxwell viscoelasticity, we
arrive at the appropriate weak form
\begin{eqnarray}
  \label{eq:EQM4}
  && \mathcal{A}(\mathbf{u},\phi \,| \, \mathbf{u}', \phi')
  - \int_{M} 2 \mu \,\mathbf{m} : \mathbf{d}' \dd^{3} \mathbf{x} + \int_{\partial M} (\mathbf{u}'\cdot \nabla \Phi + \phi' + \psi') \,\sigma \dd S \nonumber
  \\ &&+\int_{M}\rho \left(\mathbf{u}' \cdot \nabla \psi + \mathbf{u} \cdot
  \nabla \psi' \right) \dd^{3}\mathbf{x} + \bm{\omega}' \cdot \mathbf{D}\cdot
  \bm{\omega} + \int_{M} 2\mu\,(\tau \,\dot{\mathbf{m}} +
  \mathbf{m}-\mathbf{d}):\mathbf{m}' \dd^{3}\mathbf{x} = 0,
\end{eqnarray}
where an additional test function,  $\mathbf{m}'$, has been introduced
to enforce the evolution equation for the internal variable. Note that 
this equality is required to hold for all times, but with
test functions being time-independent.

Within this problem it is necessary, at each time step, to solve an
elastostatic equation forced by the load, the centrifugal potential
perturbation, and a body force associated with the internal variable.
Physically, the latter force represents the  viscoelastic relaxation
of stress. Having done this, we can compute $\mathbf{d}$ and then use the
evolution equation for $\mathbf{m}$ to advance the system in time. This is all
done subject to the initial condition
\begin{equation}
  \mathbf{m}(\mathbf{x},t_{0}) = \mathbf{0}.
\end{equation}

Because the internal variable couples to the deformation only through the
deviatoric strain, our previous discussion of uniqueness and existence carries
over immediately, with solutions of the problem defined up to the following
transformations
\begin{equation}
  \mathbf{u} \mapsto \mathbf{u} +  \mathbf{a} + \mathbf{b} \times \mathbf{x}, \quad
  \phi \mapsto \phi  - (\mathbf{a} + \mathbf{b} \times \mathbf{x})\cdot \nabla \Phi,
  \quad \bm{\omega} \mapsto \bm{\omega}, \quad \mathbf{m} \mapsto \mathbf{m},
\end{equation}
for any spatially-constant vectors  $\mathbf{a}$ and $\mathbf{b}$.

Other approaches to  viscoelasticity in the context of GIA are possible. 
 Boltzmann's superposition principle  provides a general  convolutional form for the constitutive relation 
 within  linearised viscoelasticity \citep[e.g.][]{coleman1961foundations}, 
and from this one can also arrive at the so-called correspondence principle 
using Laplace-transform methods \citep[e.g.][]{peltier1974impulse}.
Within the  application of  adjoint methods to GIA, our view is that time-domain 
formulations for viscoelasticity are most appropriate. This is because 
of their near ubiquitous use within codes for modelling GIA in laterally heterogeneous earth models.
Note, however, that in the adjoint theory of \cite{martinec2015forward}, 
the general convolutional form for linearised viscoelasticity was used.

\subsection{Gravitationally self-consistent water loads}

To complete specification of the GIA forward problem we need to link the
deformation of the earth model to changes in sea level. This leads us to the
sea level theory of \cite{farrellclark} that has been subsequently modified to
account for shoreline migration and rotational feedbacks \citep[e.g.][]{
  milnemitrovica98,mitrovicamilne,kendall}. The present discussion follows
closely that in \cite{crawford2018} and \cite{al2024reciprocity}.

The total surface load at given time is the sum of the ocean and ice loads and
can be written
\begin{equation}
  \rho_{w} \, C\, SL + \rho_{i}\,(1-C)\,I,
\end{equation}
where $\rho_{w}$ is the water density, $C$ the ocean function, $SL$ the sea level, $\rho_{i}$ the ice density,
and $I$ the ice thickness. Underlying this expression  are a host of standard definitions and assumptions
that we  briefly summarise. First, we are using a hydrostatic theory for the oceans, and hence define
the sea level to be the signed distance from the solid surface to the equipotential of gravity
 on which the  sea surface lies. To allow for
ice shelves we apply a simple hydrostatic balance, with this leading to the following definition
of the ocean function
\begin{equation}
  C(\mathbf{x},t) = \left\{
  \begin{array}{cc}
    1 & \rho_{w}\,SL(\mathbf{x},t) - \rho_{i}\ I(\mathbf{x},t) > 0   \\
    0 & \rho_{w}\,SL(\mathbf{x},t) - \rho_{i}\ I(\mathbf{x},t) \le 0 \\
  \end{array}
  \right..
\end{equation}
This function equals one in the  oceans and zero otherwise.
For later convenience, we write
\begin{equation}
  \mathcal{O}_{t} = \{ \mathbf{x} \in \partial M \,| \, \rho_{w}\,SL(\mathbf{x},t) - \rho_{i}\ I(\mathbf{x},t) > 0 \},
\end{equation}
for the subset of $\partial M$ covered by  oceans at time $t$, and
\begin{equation}
  \partial \mathcal{O}_{t} = \{ \mathbf{x} \in \partial M \,| \, \rho_{w}\,SL(\mathbf{x},t) - \rho_{i}\ I(\mathbf{x},t) = 0 \},
\end{equation}
for the associated shoreline (including the grounding line in regions with ice shelves).

Within the linearised equations of motion, it is not the total load that is
relevant but the change in the load since an initial time, $t_{0}$, when the
system is assumed be in equilibrium. We can, therefore, write the load
occurring within eq.(\ref{eq:EQM3}) as
\begin{equation}
  \label{eq:sigma}
  \sigma = \rho_{w} \, C\, SL + \rho_{i}\,(1-C)\,I - \rho_{w} \, C_{0}\, SL_{0} - \rho_{i}\,(1-C_{0})\,I_{0},
\end{equation}
where the subscript zero is used to denote values at the initial time. In terms of this load, conservation of mass between
the oceans and ice sheets can be conveniently expressed as
\begin{equation}
  \label{eq:masscon}
  \int_{\partial M } \sigma \dd S = 0.
\end{equation}
Finally, the sea level is related to the solid earth deformation through
\begin{equation}
  \label{eq:SL}
  SL = SL_{0} - \frac{1}{g}(\mathbf{u}\cdot \nabla \Phi + \phi + \psi) + \frac{\Phi_{g}}{g},
\end{equation}
where $g$ is the  acceleration due to gravity at the surface and $\Phi_{g}$ is a spatially constant field whose value
is fixed through eq.(\ref{eq:masscon}). Note that the dependence of  sea level on
the solid earth deformation is invariant under linearised rigid body motions.  Thus, the sea level is
uniquely determined from solution
of the equations of motion.  In what follows, it will sometimes be useful to
write
\begin{equation}
  SL_{1} = SL-SL_{0},
\end{equation}
for the change in sea level since the initial time.

Within eq.(\ref{eq:sigma}), the load, $\sigma$, depends on the sea level, which
is one of the variables we wish to determine. The sea level is, in turn,
related to the load-induced deformation through eq.(\ref{eq:SL}). Thus, the
forces within the problem depend on its solution, and hence we arrive at the
sea level equation of \cite{farrellclark}. In practice, this problem can be
solved efficiently using a simple iterative scheme. The assumed sea level at
each stage is used to determine a load, from this load the resulting
deformation, and hence an improved estimate for the sea level. 
Convergence of this scheme is typically obtained in only a handful of iterations.

\subsection{Summary of the forward problem}

For convenience, we  gather together the complete set of equations for the 
GIA forward problem within a unified weak form:
\begin{eqnarray}
  \label{eq:forwardfull}
 && \mathcal{A}(\mathbf{u},\phi \,| \, \mathbf{u}', \phi')
  - \int_{M} 2 \mu \mathbf{m} : \mathbf{d}' \dd^{3} \mathbf{x}
  + \int_{\partial M} \left[\mathbf{u}'\cdot \nabla \Phi +
    \phi' + \psi'\right] \sigma \dd S
   \nonumber \\ &&+\int_{M}\rho \left[\mathbf{u}'
    \cdot \nabla \psi + \mathbf{u} \cdot \nabla \psi' \right]
  \dd^{3}\mathbf{x} + \bm{\omega}' \cdot \mathbf{D}\cdot
  \bm{\omega} + \int_{M} 2\mu\,(\tau \,\dot{\mathbf{m}} +
  \mathbf{m}-\mathbf{d}):\mathbf{m}' \dd^{3}\mathbf{x}
  \nonumber \\ && + \int_{\partial M} \left[ \sigma - \rho_{w} \, C\, SL - \rho_{i}\,(1-C)\,I + \rho_{w} \,
  C_{0}\, SL_{0} + \rho_{i}\,(1-C_{0})\,I_{0}\right] g \,
  SL_{1}' \dd S \nonumber \\ &&  + \int_{\partial M}\left( g\, SL_{1} + \mathbf{u}\cdot \nabla \Phi + \phi + \psi - \Phi_{g}
  \right) \sigma' \dd S - \Phi_{g}'
  \int_{\partial M} \sigma \dd S  = 0.
\end{eqnarray}
Here new test functions, $SL_{1}'$, $\sigma'$, and $\Phi_{g}'$ associated, respectively, with eq.(\ref{eq:sigma}), eq.(\ref{eq:SL}), and eq.(\ref{eq:masscon}) have been introduced,
while we recall that the stated equality is to hold for all possible test-functions.  Note that in combining the various equations certain sign choices have been made
and scalar factors  included so as to maximise the symmetry of the resulting expression. 

Writing the forward problem in weak form is
convenient both notationally and practically. It is this form of the equations
that is necessary within the application of finite element methods
to  GIA, while the use of weak forms  simplifies the derivation and statement of
 the adjoint equations within the next section. 
If required, the corresponding
 strong form of the equations of motion can be readily obtained using integration by parts. Doing this 
would yield the usual equations for quasi-static momentum balance, 
Newtonian gravitation, and all associated boundary conditions
\citep[e.g.][]{dahlen74,trompmitrovica}.

\section{Sensitivity kernels and the first-order adjoint GIA problem}

Within this section, we apply the first-order adjoint method to obtain
sensitivity kernels for the GIA problem with respect to mantle viscosity, ice
thickness, and initial sea level (or, equivalent, palaeo-topography). 
 Later, when we discuss the second-order adjoint method,
it will be necessary to distinguish between first- and second-order adjoint
variables and equations. For the moment, however, it is understood that
in saying, for example, the  ``adjoint equations'' we mean ``first-order adjoint
equations''.

\subsection{Objective functionals and sensitivity kernels}

Consider a real-valued \emph{objective functional}
\begin{equation}
  J = J(SL,\mathbf{u}, \phi, \bm{\omega}),
\end{equation}
defined in terms of the sea level, $SL$,
surface values of $\mathbf{u}$ and $\phi$, and the perturbed angular velocity,
$\bm{\omega}$, over a time interval $\mathcal{I} = [t_{0},t_{1}]$.
Through a GIA simulation,   the arguments, $(SL,\mathbf{u}, \phi, \bm{\omega})$,
of $J$ can be calculated given suitable model parameters, and hence the value of this functional determined.
For definiteness, we will focus on the model parameters
of the widest interest within the GIA community:
\begin{enumerate}
  \item  viscosity, $\eta = \mu \tau$, defined in $M$;
  \item  ice sheet thickness, $I$, defined on $\partial M \times \mathcal{I}$;
  \item  initial sea level, $SL_{0}$, at time $t_{0}$, defined on $\partial M$.
\end{enumerate}
Other properties of the Earth such as  elastic modulii, density, or topography on internal
boundaries might also be considered. Such quantities are, however, best constrained
through seismological studies and are likely to have comparatively little impact
in the context of GIA \citep[e.g.][]{austermann2021effect}. We can now define a \emph{reduced objective functional} through
\begin{equation}
  \hat{J}(\eta, I , SL_{0}) = J(SL,\mathbf{u}, \phi, \bm{\omega}),
\end{equation}
where it is understood that the arguments on the right-hand side have
been determined from those on the left through the solution of the GIA
forward problem.

If the model parameters are perturbed, the first-variation of
$\hat{J}$ can be written as
\begin{equation}
  \delta \hat{J} = \int_{M} D_{\eta} \hat{J}    \, \delta \eta \dd^{3} \mathbf{x}
  + \int_{\mathcal{I}} \int_{\partial M} D_{I} \hat{J} \, \delta I \dd S \dd t + \int_{\partial M} D_{SL_{0}} \hat{J} \,\delta SL_{0} \dd S,
\end{equation}
with the  functional derivatives $D_{\eta}\hat{J} $, $D_{I}\hat{J} $,
and $D_{SL_{0}}\hat{J} $ defined through this expression. These
 functional derivatives are equivalently called  \emph{sensitivity kernels},
 and were denoted by
 $K_{\eta}$, $K_{I}$, and $K_{SL_{0}}$ within  \cite{crawford2018}.

Recall that solutions of the GIA forward problem are only unique up to a
time-dependent linearised rigid body motion. In order for the objective
functional to be well-defined we require that $J$ be invariant with respect to
such transformations, this meaning
\begin{equation}
  \label{eq:Jinv}
  J(SL,\mathbf{u} + \mathbf{a} + \mathbf{b}\times \mathbf{x}, \phi
  - (\mathbf{a} + \mathbf{b}\times \mathbf{x})\cdot \nabla \Phi, \bm{\omega})
  = J(SL,\mathbf{u}, \phi, \bm{\omega}),
\end{equation}
for all spatially-constant vectors $\mathbf{a}$ and $\mathbf{b}$. As an example,
any $J$ defined solely in terms of the sea level satisfies this condition. This
is because sea level is the \emph{relative} distance between two surfaces. Conversely,
the value of the radial displacement at a given point and time provides an example of an invalid  objective functional. 

\subsection{PDE constrained optimisation via the Lagrangian method}

The first-variation of $\delta J$ with respect to $(SL,\mathbf{u}, \phi,
  \bm{\omega})$ can be written
\begin{equation}
  \label{eq:dJdSL}
  \delta J = \int_{\mathcal{I}} \left\{ \int_{\partial M} \left( D_{SL}J\,\delta SL + D_{\mathbf{u}}J \cdot \delta \mathbf{u}
  +D_{\phi}J \, \delta
  \phi \right) \dd S + D_{\bm{\omega}}J\cdot \delta
  \bm{\omega} \right\}\!\dd t,
\end{equation}
for appropriate functional derivatives. The invariance condition for the objective
functional in eq.(\ref{eq:Jinv}) implies that
\begin{equation}
  \label{eq:dJinv}
  \int_{\partial M} \left( D_{\mathbf{u}}J - D_{\phi}J\, \nabla \Phi \right) \cdot(\mathbf{a} + \mathbf{b}\times \mathbf{x}) \dd S
  = 0,
\end{equation}
with $\mathbf{a}$ and $\mathbf{b}$ arbitrary constant vectors.

To determine the variation of $\hat{J}$ with respect to the model parameters,
$(\eta, I , SL_{0})$, we apply the Lagrangian method for PDE constrained
optimisation \citep[e.g.][Section 2.10]{troltzsch}. At the outset, it will be
useful to introduce a \emph{time-reversal operator}, $\mathcal{T}$. Let $f$
be a function defined on the time-interval $\mathcal{I} = [t_{0},t_{1}]$, and
set
\begin{equation}
  (\mathcal{T}f)(t) = f(t_{1} - t + t_{0}).
\end{equation}
This definition applies identically to scalar-, vector-, and tensor-valued functions
and to those that depend on position also.  It is clear that
$\mathcal{T}$ is involutary (i.e., $\mathcal{T}^{-1} = \mathcal{T}$), that
it anti-commutes with time-differentiation, meaning
\begin{equation}
  \frac{\ddns}{\ddns t} (\mathcal{T}f) = - \mathcal{T}\left(\frac{\ddns f}{\ddns t}\right),
\end{equation}
and that it is self-adjoint relative to the standard inner product on $L^{2}(\mathcal{I})$.
In the case of scalar-valued functions, the latter condition implies
\begin{equation}
  \int_{\mathcal{I}} (\mathcal{T}f) \, g \dd t = \int_{\mathcal{I}} f \, (\mathcal{T}g) \dd t,
\end{equation}
for arbitrary $f$ and $g$, while the extension to vector- and tensor-valued functions
is obvious. We now define a Lagrangian for the problem
\begin{eqnarray}
  \label{eq:Lag1}
  L = J + \int_{\mathcal{I}} \!\!\!\!\!\!&&\!\!\!\!\!\! \left\{
  \mathcal{A}(\mathbf{u},\phi \,| \, \mathcal{T} \mathbf{u}^{\dagger}, \mathcal{T}\phi^{\dagger})
  - \int_{M} 2 \mu \mathbf{m} : (\mathcal{T}\mathbf{d}^{\dagger}) \dd^{3} \mathbf{x}
  + \int_{\partial M} \left[(\mathcal{T}\mathbf{u}^{\dagger})\cdot \nabla \Phi +
    \mathcal{T}\phi^{\dagger} + \mathcal{T}\psi^{\dagger}\right] \sigma \dd S
  \right. \nonumber \\ &&+\int_{M}\rho \left[(\mathcal{T}\mathbf{u}^{\dagger})
    \cdot \nabla \psi + \mathbf{u} \cdot \nabla (\mathcal{T}\psi^{\dagger}) \right]
  \dd^{3}\mathbf{x} + (\mathcal{T}\bm{\omega}^{\dagger}) \cdot \mathbf{D}\cdot
  \bm{\omega} + \int_{M} 2\mu\,(\tau \,\dot{\mathbf{m}} +
  \mathbf{m}-\mathbf{d}):(\mathcal{T}\mathbf{m}^{\dagger}) \dd^{3}\mathbf{x}
  \nonumber \\ && + \int_{\partial M} \left[ \sigma - \rho_{w} \, C\, SL - \rho_{i}\,(1-C)\,I + \rho_{w} \,
  C_{0}\, SL_{0} + \rho_{i}\,(1-C_{0})\,I_{0}\right] g \,
  (\mathcal{T}SL_{1}^{\dagger}) \dd S \nonumber \\ && \left. + \int_{\partial M}\left( g\, SL_{1} + \mathbf{u}\cdot \nabla \Phi + \phi + \psi - \Phi_{g}
  \right) (\mathcal{T}\sigma^{\dagger}) \dd S - (\mathcal{T}\Phi_{g}^{\dagger})
  \int_{\partial M} \sigma \dd S \right\} \dd t.
\end{eqnarray}
Here, $J$ is the objective functional of interest, while the remaining terms
are associated with the constraint that the \emph{state variables}, ($\mathbf{u}$, $\phi$,
$\bm{\omega}$, $\mathbf{m}$, $SL$, $\sigma$, $\Phi_{g}$), satisfy the
forward problem. For each state variable there is a corresponding
\emph{adjoint state variable} indicated by the same symbol but with the addition
of a dagger superscript. It is worth emphasising that within this definition  the state and adjoint state variables
are independent of one another, with the dagger notation simply providing a labelling scheme for the new variables.
  Time reversal of the adjoint state variables
within the Lagrangian is not necessary but has been done for later convenience.

The Lagrange multiplier theorem shows that if $L$ is stationary with respect to
both the state variables and the adjoint state variables then functional
derivatives of $L$ and $\hat{J}$ with respect to the model parameters are
equal. By construction, $L$ being stationary with respect to the adjoint state
variables implies that the state variables solve the forward problem. Indeed, 
taking the necessary variation and requiring the equality hold for 
each time separately, we arrive at the unified weak form 
of the GIA problem stated within eq.(\ref{eq:forwardfull}).

Requiring
that $L$ is stationary with respect to the state variables, on the other hand,
defines a new system of equations that must be satisfied by the adjoint state
variables. To obtain these latter equations, we first set the variation of $L$
with respect to $(\mathbf{u},\phi,\bm{\omega},\mathbf{m})$ equal to zero. Using
eq.(\ref{eq:dJdSL}), this implies
\begin{eqnarray}
  \int_{\mathcal{I}} \!\!\!\!\!\!&&\!\!\!\!\!\! \left\{ \int_{\partial M} \left(D_{\mathbf{u}}J\cdot \delta \mathbf{u} +
  D_{\phi}J\,\delta \phi\right) \dd S + D_{\bm{\omega}}J \cdot \delta \bm{\omega} + \mathcal{A}(\delta \mathbf{u},\delta
  \phi \,| \, \mathcal{T}\mathbf{u}^{\dagger}, \mathcal{T}\phi^{\dagger}) -
  \int_{M} 2 \mu \delta \mathbf{m} : (\mathcal{T}\mathbf{d}^{\dagger}) \dd^{3}
  \mathbf{x} \right. \nonumber \\ && +\int_{M}\rho
  \left[(\mathcal{T}\mathbf{u}^{\dagger}) \cdot \nabla \delta\psi + \delta
    \mathbf{u} \cdot \nabla (\mathcal{T}\psi^{\dagger}) \right] \dd^{3}\mathbf{x} +
  (\mathcal{T}\bm{\omega}^{\dagger}) \cdot \mathbf{D}\cdot \delta \bm{\omega} +
  \int_{M} 2\mu\,(\tau \,\dot{\delta\mathbf{m}} + \delta \mathbf{m}-\delta
  \mathbf{d}):(\mathcal{T}\mathbf{m}^{\dagger}) \dd^{3}\mathbf{x} \nonumber \\ &&
  \left. +\int_{\partial M}\left( \delta \mathbf{u}\cdot \nabla \Phi + \delta \phi + \delta \psi
  \right) (\mathcal{T}\sigma^{\dagger}) \dd S \right\} \dd t = 0.
\end{eqnarray}
Integrating by parts with respect to time and
using the properties of $\mathcal{A}$, $\mathbf{D}$, and  $\mathcal{T}$, we find that the
above equality reduces to
\begin{eqnarray}
  && \mathcal{A}(\mathbf{u}^{\dagger},\phi^{\dagger}\,|\, \delta \mathbf{u}, \delta \phi) -
  \int_{M} 2\mu \, \mathbf{m}^{\dagger}: \delta \mathbf{d} \dd^{3}\mathbf{x}
  + \int_{\partial M}\left( \delta \mathbf{u}\cdot \nabla \Phi + \delta \phi + \delta \psi
  \right) \sigma^{\dagger} \dd S +\int_{\partial M} \left[\left(\mathcal{T} D_{\mathbf{u}}J \right)\cdot
    \delta \mathbf{u} + \left(\mathcal{T} D_{\phi}J\right)\delta \phi\right]\! \dd S \nonumber \\ && + \int_{M}\rho\,(
  \mathbf{u}^{\dagger}\cdot \nabla \delta \psi + \delta \mathbf{u}\cdot \nabla
  \psi^{\dagger}) \dd^{3}\mathbf{x} + \delta \bm{\omega} \cdot
  \left[\mathbf{D}\cdot \bm{\omega}^{\dagger} + \left(\mathcal{T}D_{\bm{\omega}}J \right)\right] + \int_{M}2\mu \left( \tau
  \dot{\mathbf{m}}^{\dagger} + \mathbf{m}^{\dagger} -
  \mathbf{d}^{\dagger}\right):\delta \mathbf{m} \dd^{3}\mathbf{x} = 0,
\end{eqnarray}
where $(\delta\mathbf{u},
  \delta\phi, \delta \bm{\omega}, \delta \mathbf{m})$ are now arbitrary time-independent
test functions, along with the initial condition
\begin{equation}
  \mathbf{m}^{\dagger}(\mathbf{x},t_{0}) = \mathbf{0}.
\end{equation}

Variation of $L$ with respect to $\sigma$ and $\Phi_{g}$ are trivial, yielding
the following relations
\begin{equation}
  SL_{1}^{\dagger} = -\frac{1}{g}(\mathbf{u}^{\dagger}\cdot \nabla \Phi + \phi^{\dagger} + \psi^{\dagger}) + \frac{\Phi_{g}^{\dagger}}{g}, \quad
  \int_{\partial M} \sigma^{\dagger} \dd S = 0.
\end{equation}
In varying the Lagrangian with respect to sea level, we
first note that $SL_{0}$ is a parameter in the problem, and so is
fixed within this variation. This means that it is not $SL$ but the sea level
change, $SL_{1} = SL-SL_{0}$, that undergoes arbitrary variation (it is for this reason we called the corresponding
adjoint variable $SL_{1}^{\dagger}$). The other complication is
that the ocean function depends on the sea level, and hence must be varied as
part of the calculation. To do this, we recall the following expression
\begin{equation}
  \delta C = \frac{\rho_{w}\delta SL_{1}}{\rho_{w} \partial_{\perp} SL - \rho_{i}
    \partial_{\perp} I} \delta_{\partial \mathcal{O}_{t}},
\end{equation}
for the first variation of the ocean function with respect to a change in sea level
\citep[][eq.(A30)]{crawford2018}. Here $\partial_{\perp}$ denotes the directional derivative perpendicular to the shoreline
oriented towards the oceans, and $\delta_{\partial \mathcal{O}_{t}}$ is the Dirac line distribution on the shoreline. The
action of this distribution on a smooth test function, $f$, is given by
\begin{equation}
  \int_{\partial M} f\, \delta_{\partial \mathcal{O}_{t}} \dd S = \int_{\partial \mathcal{O}_{t}} f \dd l,
\end{equation}
with $\dd l$ the standard line element. A key simplification within the derivation is
that the singular term $\delta C$ is multiplied by the smooth function $\rho_{w} SL - \rho_{i}
  I$ which, by definition, vanishes on $\partial \mathcal{O}_{t}$. The remainder of the calculation is easy, with the
result being
\begin{equation}
  \sigma^{\dagger} = \rho_{w}\, C^{\dagger}\, SL^{\dagger}_{1} - \frac{1}{g} \mathcal{T}D_{SL}J,
\end{equation}
where we have set $C^{\dagger} = \mathcal{T} C$ for convenience.

The existence and uniqueness for solutions for the adjoint problem can be
readily analysed. First, it is clear that solutions, $(\mathbf{u}^{\dagger},
  \phi^{\dagger})$ of the elastostatic sub-problem are defined only up to a
linearised rigid body motion, but that this non-uniqueness does not propagate
through to the other adjoint variables. The existence condition for the adjoint
problem can be obtained by taking the test functions, $(\delta
  \mathbf{u},\delta \phi)$, to be a linearised rigid body motion, this leading to
the equality
\begin{equation}
  \int_{\partial M} \left( D_{\mathbf{u}}J - D_{\phi}J\,\nabla \Phi \right) \cdot(\mathbf{a} + \mathbf{b}\times \mathbf{x}) \dd S
  + \int_{M}\rho\, (\mathbf{a} + \mathbf{b}\times \mathbf{x})\cdot \nabla
  \psi^{\dagger} \dd^{3}\mathbf{x} = 0,
\end{equation}
for arbitrary constant vectors $\mathbf{a}$ and $\mathbf{b}$.
The first integral vanishes because of the assumed invariance of $J$, while we have seen
that the second integral must be zero in order for the forward problem to have solutions.
Thus, so long as the objective functional is physically well defined the
adjoint problem admits solutions.

\subsection{Practical implementation}

\label{sec:adjoint_prac}

The adjoint GIA problem is nearly identical in form to the forward problem.
This is useful because it means that a code that can solve the forward problem
can also solve the adjoint problem with minimal modifications. There are just
two areas of difference. First, the adjoint problem involves more general force
terms. In the forward problem, there is only the applied ice load. But in the
adjoint problem there are four force terms, each being expressed in terms of a
functional derivative of $J$ with respect to one of its arguments. The
term associated with $D_{SL}J$ plays a role similar to the ice load within
the forward problem, while  the term due to  $D_{\bm{\omega}}J$ acts like an
 angular momentum perturbation when solving for $\bm{\omega}^{\dagger}$. Finally, 
the   force terms associated with $D_{\mathbf{u}}J$
and $D_{\phi}J$ correspond to inhomogeneous boundary conditions
on the surface
of the earth model. For the convenience of the reader, we now summarise these
boundary conditions explicitly. To put these results in context, we first
recall that within the forward GIA problem the relevant boundary conditions
for $\mathbf{u}$ and $\phi$ on $\partial M$ are
\begin{eqnarray}
  &&\unvec\cdot \left(
  \kappa \nabla \cdot \mathbf{u}  + 2\mu \mathbf{d}
  \right) = -\sigma \nabla \Phi +  \unvec \cdot  2\mu \mathbf{m},  \\
  && \diff{(4\pi G)^{-1} \unvec \cdot \nabla \phi} - \rho \unvec\cdot \mathbf{u} = \sigma, 
\end{eqnarray}
where $\diff{\cdot}$ denotes a jump in a quantity across $\partial M$ in the direction of the outward unit normal.
Within the adjoint problem, these conditions are generalised to
\begin{eqnarray}
  &&\unvec\cdot \left(
  \kappa \nabla \cdot \mathbf{u}^{\dagger}  + 2\mu \mathbf{d}^{\dagger}
  \right) = -\sigma^{\dagger} \nabla \Phi +  \unvec \cdot 2 \mu \mathbf{m}^{\dagger} - \mathcal{T} D_{\mathbf{u}}J,  \\
  && \diff{(4\pi G)^{-1} \unvec \cdot \nabla \phi^{\dagger}} - \rho \unvec\cdot \mathbf{u}^{\dagger} = \sigma^{\dagger}
  + \mathcal{T} D_{\phi} J.
\end{eqnarray}
These two additional terms within the surface boundary conditions are the \emph{only} differences between the
forward and adjoint elastostatic sub-problems, and hence a code that can solve the forward problem
can be readily adapted to solve the adjoint one.

The remaining difference between the forward and adjoint GIA problems  is that the ocean function, $C^{\dagger}$, occurring
within the adjoint problem is not determined dynamically. Instead, it is the
time-reversed ocean function obtained from solution of the forward problem.
This means that the adjoint problem is linear but with time-dependent
coefficients.

If shoreline migration is neglected within the forward problem,
then the load reduces to
\begin{equation}
  \sigma = \rho_{w} \, C\, SL_{1} + \rho_{i}\,(1-C)\,I_{1},
\end{equation}
where $SL_{1}$ and $I_{1}$ denote, respectively, the change in sea level  and ice thickness
since $t_{0}$. The forward problem is then linear, and the adjoint equations differ
from the forward problem only through the inclusion of more general force terms.

\subsection{Expressions for the sensitivity kernels}

We can now derive expressions for the sensitivity kernels, $D_{\eta}\hat{J} $, 
$D_{I}\hat{J} $, and $D_{SL_{0}} \hat{J}$ by varying the Lagrangian with respect to the model
parameters. The first variation of $\hat{J}$ with respect to $\eta$ is given by
\begin{equation}
  \delta \hat{J} = \int_{\mathcal{I}} \int_{M} 2 \delta \eta \,
  \dot{\mathbf{m}}:(\mathcal{T}\mathbf{m}^{\dagger}) \dd^{3}\mathbf{x} \dd t,
\end{equation}
where we have used $\eta = \mu \, \tau$ and recall that the shear modulus, $\mu$, is fixed.
Using the evolution equation for $\mathbf{m}$, it follows that the viscosity kernel can be written as
\begin{equation}
  \label{eq:Keta}
  D_{\eta}\hat{J}= \int_{\mathcal{I}} \frac{2\mu }{\eta} \,
  (\mathbf{d}-\mathbf{m}):(\mathcal{T}\mathbf{m}^{\dagger})  \dd t.
\end{equation}

Turning to the kernel for ice thickness, we need to account for changes in the
ocean function when taking variations of the Lagrangian. To do this, we again
apply eq.(A30) of \cite{crawford2018} which, in this instance, takes the form
\begin{equation}
  \delta C = \frac{-\rho_{w}\delta I}{\rho_{w} \partial_{\perp} SL - \rho_{i}
    \partial_{\perp} I} \delta_{\partial \mathcal{O}_{t}}.
\end{equation}
As previously, singular terms on the shoreline cancel, and
we are left with
\begin{equation}
  \delta \hat{J} =  -\int_{\mathcal{I}}
  \int_{\partial M} \rho_{i} \, g \left[ (1-C) \delta I - (1-C_{0}) \delta I_{0} \right]
  (\mathcal{T}SL_{1}^{\dagger}) \dd S \dd t.
\end{equation}
Using a Dirac delta function,  we can identify the ice kernel as
\begin{equation}
  \label{eq:Kice}
  D_{I}\hat{J} = - \rho_{i} \, g \left[(1-C) \, (\mathcal{T}SL_{1}^{\dagger})
  -  (1-C_{0})\int_{\mathcal{I}} \mathcal{T} SL_{1}^{\dagger} \dd t \,\delta(t-t_{0})
  \right].
\end{equation}
The derivation of the kernel for initial sea level proceeds similarly, leading to the expression
\begin{equation}
  \label{eq:KSL0}
  D_{SL_{0}}\hat{J} = \int_{\mathcal{I}} \left[  D_{SL}J - \rho_{w} g \,(C-C_{0})
  (\mathcal{T}SL_{1}^{\dagger}) \right] \dd t.
\end{equation}

The sensitivity kernels are expressed as combinations of the forward and adjoint state variables.
We know the displacement vector and gravitational
potential perturbation within the forward and adjoint problems are only defined up to a linearised
rigid body motion. From the above expressions it can, however, be readily seen that the kernels are
invariant under such transformations, and hence  uniquely defined.

\subsection{An alternate form for the viscosity kernel}

Within this paper, we have made use of an internal variable
method to account for the Maxwell rheology. While this approach is widely used
  and easy to
implement \citep[e.g.][]{simohughes,nsm},  a range of 
 equivalent methods have been considered within
the GIA literature. For the convenience of a reader hoping to implement 
the adjoint method within their own codes, we now present an alternative
expression for the
viscosity kernel which is independent of the internal variable formalism.

First, we recall that the stress, $\mathbf{T}$, takes the form
\begin{equation}
  \mathbf{T} = \kappa \,\nabla \cdot \mathbf{u}\, \mathbf{1} + 2\mu \,(\mathbf{d} - \mathbf{m}),
\end{equation}
and hence the deviatoric component of the stress, $\bm{\tau}$, can be expressed as 
\begin{equation}
  \bm{\tau} = 2\mu \,(\mathbf{d} - \mathbf{m}).
\end{equation}
The deviatoric stress can then be further split as
\begin{equation}
  \bm{\tau} = \bm{\tau}_{e} + \bm{\tau}_{v}, 
\end{equation}
with $\bm{\tau}_{e} = 2\mu\, \mathbf{d}$ the elastic part and  $\bm{\tau}_{v} = \bm{\tau}-\bm{\tau}_{e}$
the viscous. Such  a decomposition  is present within all other  time-domain approaches
to viscoelasticity with the GIA literature \citep[e.g.][]{hanyk95,zhong,latychev,bailey2006large}.
Applying the same decomposition to the stress within the adjoint problem, we can then  write
eq.(\ref{eq:Keta}) as
\begin{equation}
  \label{eq:KetaAlt}
  D_{\eta}\hat{J}= -\int_{\mathcal{I}} \frac{1}{2\mu \eta} \bm{\tau} :(\mathcal{T}\bm{\tau}_{v}^{\dagger}) \dd t.
\end{equation}
In this way, we  see that the calculation of sensitivity kernels using the 
adjoint method does not depend on the use of internal variables for
 modelling the Maxwell rheology. Rather, any numerical code
for modelling GIA is, subject to the minor modifications to the force terms
discussed above, 
capable of determining all variables needed to form the sensitivity kernels.

\subsection{Comparison with related studies}

The adjoint equations and sensitivity kernels derived within this 
section are new, but they build on and extend results from earlier studies
 \citep[e.g.][]{alattartromp,martinec2015forward,  crawford2018,al2024reciprocity}.
 In particular, \cite{crawford2018} derived   adjoint equations 
 and sensitivity kernels for GIA within a Maxwell earth model
  in the absence of rotational feedbacks.
 Crawford's results were numerically implemented, with the forward 
 calculations benchmarked against another GIA code, and 
  tests performed to verify that the sensitivity
 kernels correctly predicted the linearised dependence
 of chosen objective functionals on the underlying model parameters.
 Subsequent work by \cite{crawford2019viscoelastic} and \cite{lloyd2023}  extended these
 numerical tests further by considering sensitivity kernels calculated relative to
 laterally heterogeneous viscosity models.

 If terms associated with rotational feedbacks are removed, then the 
 results of this paper should be equivalent to those within 
 \cite{crawford2018}. Verifying this correspondence is 
 complicated due the two papers being based on different, 
 but equivalent, formulations of the GIA forward problem. 
 Moreover, the adjoint variables in each paper are defined
 relative to their respective forward problems. This means, for example, 
 that the adjoint sea level in this 
 paper is not equal to the adjoint sea level within 
 \cite{crawford2018}. Nevertheless, through a lengthy
 but simple algebraic process the 
 required equivalence of the results can be established.
 
 Beyond the direct calculation just mentioned, the consistency 
 of our results with those of \cite{crawford2018} 
 can be seen in a number of ways. We noted in Section \ref{sec:adjoint_prac}  that 
 if shoreline migration is neglected, then the forward 
 and adjoint problems have 
  the same form except for the adjoint 
 equations involving more general force terms. 
 Precisely the same result was found to hold within \cite{crawford2018}
 relative to their formulation of the forward GIA problem. 
 Similarly, when shoreline migration is included within either approach, the  adjoint problems 
 involve the time-reversed ocean function obtained 
 through solution of the forward problem. 

 A further useful point of comparison is with \cite{al2024reciprocity}. That paper was
  concerned with the calculation of  sensitivity kernels for the 
 elastic sea level fingerprint problem incorporating rotational feedbacks, and 
 contains detailed numerical checks of the theoretical results. The elastic
 fingerprint problem is a special case of the GIA problem 
 in which viscoelasticity and shoreline migration are neglected. On making these
 approximations, it is readily seen that the results of this paper reduce 
 precisely to those in \cite{al2024reciprocity}.

\subsection{Singular solutions of the adjoint equations}

Within this section, we show that solutions of the adjoint equations possess
singularities as functions of time whenever the objective functional depends on
the state variables at a discrete set of times. Such behaviour cannot be
dismissed as a mere pathology, but occurs frequently in practice. Suppose for
simplicity that the objective functional, $J$, depends on the state variables
at a single observation time, $t'$. The functional derivatives of $J$ then take
the form
\begin{eqnarray}
  D_{SL}J \!\!\!\!&=&\!\!\!\!
  \tilde{h}_{SL}^{\dagger}(\mathbf{x})\, \delta (t- t'), \\
  D_{\mathbf{u}}J
  &=& \tilde{ \mathbf{h}}_{\mathbf{u}}^{\dagger}(\mathbf{x})\, \delta (t- t'), \\
  D_{\phi}J
  &=& \tilde{h}_{\phi}^{\dagger}(\mathbf{x})\, \delta (t- t'), \\
  D_{\bm{\omega}}J
  &=& \tilde{ \mathbf{h}}_{\bm{\omega}}^{\dagger}\, \delta (t- t'),
\end{eqnarray}
for some $(\tilde{h}_{SL}^{\dagger}, \tilde{ \mathbf{h}}_{\mathbf{u}}^{\dagger}, \tilde{h}_{\phi}^{\dagger},
  \tilde{ \mathbf{h}}_{\bm{\omega}}^{\dagger} )$. To see this, note that, for example, we have 
  \begin{equation}
      \int_{\mathcal{I}} \int_{\partial M} D_{SL} J \, \delta SL \dd S \dd t = \int_{\partial M} \tilde{h}_{SL}^{\dagger}(\mathbf{x}) \, \delta SL(\mathbf{x},t') \dd S, 
  \end{equation}
  with the term on the right hand side depending on the sea level perturbation at the required observation time.
Prior to the reversed observation time, $t'' = t_{1} -t'+t_{0}$, it is clear that all adjoint
state variables vanish. If we integrate the adjoint equations  over the interval
$(t''-\epsilon, t''+\epsilon)$ and take the limit $\epsilon \rightarrow 0$ we obtain
\begin{eqnarray}
  && \mathcal{A}(\tilde{\mathbf{u}}^{\dagger},\tilde{\phi}^{\dagger}\,|\, \delta \mathbf{u}, \delta \phi)
  + \int_{\partial M}\left( \delta \mathbf{u}\cdot \nabla \Phi + \delta \phi + \delta \psi
  \right) \tilde{\sigma}^{\dagger} \dd S +\int_{\partial M} (\tilde{\mathbf{h}}^{\dagger}_{\mathbf{u}} \cdot \delta \mathbf{u} +
  \tilde{h}^{\dagger}_{\phi}\,\delta \phi) \dd S \nonumber \\ && +
  \int_{M}\rho\,( \tilde{\mathbf{u}}^{\dagger}\cdot \nabla \delta \psi + \delta
  \mathbf{u}\cdot \nabla \tilde{\psi}^{\dagger}) \dd^{3}\mathbf{x} + \delta
  \bm{\omega} \cdot \left(\mathbf{D}\cdot \tilde{\bm{\omega}}^{\dagger} +
  \tilde{\mathbf{h}}^{\dagger}_{\bm{\omega}}\right) + \int_{M}2\mu \left( \tau
  \diff{\mathbf{m}^{\dagger}} - \tilde{\mathbf{d}}^{\dagger}\right):\delta
  \mathbf{m} \dd^{3}\mathbf{x} = 0, \\ && \tilde{SL}_{1}^{\dagger} =
  -\frac{1}{g}(\tilde{\mathbf{u}}^{\dagger}\cdot \nabla \Phi +
  \tilde{\phi}^{\dagger} + \tilde{\psi}^{\dagger}) +
  \frac{\tilde{\Phi}_{g}^{\dagger}}{g}, \\ && \tilde{\sigma}^{\dagger} =
  \rho_{w}\, C^{\dagger}\, \tilde{SL}^{\dagger}_{1} -\frac{1}{g}
  \tilde{h}^{\dagger}_{SL}, \\ && \int_{\partial M} \tilde{\sigma}^{\dagger} \dd S = 0,
\end{eqnarray}
where for a time-dependent variable, $f$, we have introduced notations
\begin{equation}
  \tilde{f} =\lim_{\epsilon\rightarrow 0}
  \int_{t''-\epsilon}^{t''+\epsilon} f \dd t, \quad
  \diff{f} = \lim_{\epsilon\rightarrow 0}[f(t'' + \epsilon) - f(t''-\epsilon)],
\end{equation}
noting that $\tilde{f}$ vanishes for an integrable function. 
It follows that, with the exception  of $\mathbf{m}^{\dagger}$,
the adjoint state variables have delta-function singularities
at $t''$, with their relative amplitudes determinable through solution of a  \emph{generalised
  fingerprint problem} of the form discussed in \cite{al2024reciprocity}. Having
done this, the finite-jump in the adjoint internal variable is given by
\begin{equation}
  \diff{ \mathbf{m}^{\dagger}} = \frac{1}{\tau} \tilde{\mathbf{d}}^{\dagger}.
\end{equation}
For times $t > t''$, we can
use $\diff{\mathbf{m}^{\dagger}}$ as an initial condition to
integrate the evolution equation for $\mathbf{m}^{\dagger}$, obtaining smooth solutions
for all adjoint variables within the interval $(t'',t_{1}]$.
By linearity of the adjoint equations, these arguments  extend trivially to cases
where the objective functional depends on the state variables
at a finite number of observation  times.

These results show that between observation times the adjoint GIA problem can
be time-stepped using standard methods. At each observation time, however, it
is necessary to solve a generalised fingerprint problem whose solution
determines a finite-jump in $\mathbf{m}^{\dagger}$ along with singular
contributions to the other adjoint state variables. For the viscosity kernel in
eq.(\ref{eq:Keta}), all terms within the integrand are bounded and hence the
integral can be evaluated using a standard quadrature scheme applied within
each sub-interval between the observation times. In the case of the initial sea
level kernel in eq.(\ref{eq:Kice}), both $D_{SL}J$ and
the adjoint sea level lead to delta-function singularities at the observation
times. Using the defining property of the delta-function, each singularity
makes a discrete contribution to the integral whose value is determined through
solution of the generalised fingerprint problem, while between the observation
times the integrand is smooth and a standard quadrature scheme can be applied.

\subsection{Singular ice kernel and gradient-based optimisation}

Singularities within the solution of the adjoint GIA problem have no effect on the 
use of sensitivity kernels for viscosity and initial sea level, and complicate 
their  calculation only slightly. In the case of the ice thickness kernel,
\begin{equation}
  D_{I}\hat{J} = - \rho_{i} \, g \left[(1-C) \, (\mathcal{T}SL_{1}^{\dagger})
  -  (1-C_{0})\int_{\mathcal{I}} \mathcal{T} SL_{1}^{\dagger} \dd t \,\delta(t-t_{0})
  \right],
\end{equation}
the situation is more involved, with this kernel
having delta-function singularities at the initial time and each observation time. Within
the context of gradient-based optimisation, it is typical for
descent directions to be formed from linear combinations of the sensitivity kernel
at the current and previous iterations. But we clearly cannot
add  these singular ice kernels to an ice thickness
model.   Before outlining   methods for addressing this problem, it is worth
emphasising that the singularities of the ice kernel are real features  reflecting the
physics of GIA. Within this quasi-static theory, the instantaneous application of
a load at time, $t'$, produces an instantaneous elastic
response, and hence a discontinuous solution of the forward GIA problem.
It follows that the linearised dependence of the state variables  at  time, $t'$,
can be decomposed into the sum of two terms, the first associated with elastic deformation
at the observation time,
and the second  due to viscoelastic relaxation from earlier loading. It is
the former contribution that produces the delta-function singularity within the ice thickness
kernel.

The simplest method for working with the singular ice kernel is to parameterise
the ice thickness using a finite dimensional set of continuous basis functions.
We then need only consider the projection of the ice kernel along each basis
function, this requiring integration of the singular kernel against continuous
functions. The introduction of such a model parameterisation is, however,
necessarily ad hoc. Moreover, as the size of
the basis set is increased, the projection of these singular kernels cannot
converge point-wise, and hence non-physical ringing and other artefacts will  be introduced.

A better method was presented by \cite{alattartromp} and \cite{crawford2018}
using the idea of \cite{backus1970inferenceIII} for \emph{quelling} singular
kernels. In fact, it was this idea that  motivated the introduction
of the rate-formulation of the GIA problem, with the quelling process
there being built directly into the description of the forward problem.
The idea can be applied more generally, however. The  key step is to not regard the ice
thickness  as a model parameter, but to define it implicitly through its initial
value, $I_{0}$, and its time-derivative, $\dot{I}$. To proceed, we recall that
the first-order change in $\hat{J}$ is given by
\begin{equation}
  \delta \hat{J} =  -\int_{\mathcal{I}} \int_{\partial M} \rho_{i} g \left[ (1-C) \delta I - (1-C_{0}) \delta I_{0} \right]
  (\mathcal{T} SL_{1}^{\dagger}) \dd S \dd t,
\end{equation}
and seek new kernels, $D_{I_{0}}\hat{J}$ and $D_{\dot{I}}\hat{J}$, such that we can equivalently write
\begin{equation}
  \delta \hat{J} =   \int_{\partial M} D_{I_{0}}\hat{J}\,\delta I_{0}\dd S +
  \int_{\mathcal{I}} \int_{\partial M} D_{\dot{I}}\hat{J} \,\delta \dot{I} \dd S \dd t.
\end{equation}
Requiring equality for arbitrary $\delta I$,  integrating by parts,
and solving a  trivial ODE, we find
\begin{eqnarray}
  D_{\dot{I}}\hat{J}(\mathbf{x},t) =  -\rho_{i} g \int_{t}^{t_{1}}  \, [1-C(\mathbf{x},s)]\,  SL_{1}^{\dagger}(\mathbf{x},t_{1} - s + t_{0}) \dd s, \quad
  D_{I_{0}}\hat{J}(\mathbf{x}) = \rho_{i} g \int_{\mathcal{I}} [C(\mathbf{x},t)-C_{0}(\mathbf{x})] \,SL_{1}^{\dagger}(\mathbf{x},
  t_{1}-t+t_{0}) \dd t.
\end{eqnarray}
By inspection,
$D_{\dot{I}}\hat{J}$ undergoes only finite jumps at the observation times,
and hence it is sufficiently regular for use within standard gradient-based
optimisation schemes. Descent directions for $I_{0}$ and $\dot{I}$
can, therefore, be obtained in a normal manner, and through a
further time-integration they define an update for
the ice thickness that is continuous in time.

A third, and likely best, approach is suggested by the recent work of
\cite{zuberi2017mitigating} and \cite{syvret2022theoretical} in seismic
tomography. The application of this method to GIA will be discussed in detail
elsewhere. For the moment, we just note that it has a rigorous foundation in
functional analysis based on the Sobolev
embedding theorem \citep[e.g.][]{treves1975basic}, and allows for stronger regularity requirements to be
imposed on the ice thickness (and other model parameters) in both time and
space.

\subsection{Constraining the initial sea level}

Within the GIA forward problem, the initial sea level has been regarded as a
parameter. In practice, however, this value is usually constrained by requiring
that the calculated present-day sea level matches the observed value. Let $t_{p}\le t_{1}$
denote the time of the present and $SL_{p}$ the observed sea level. For a given
initial sea level, $SL_{0}$, we can compute $SL(t_{p})$ through solution of the
GIA forward problem. In general, this will not agree with $SL_{p}$,
but a simple iterative procedure can be applied that converges rapidly \citep[e.g.][]{johnston1993effect,peltier1994ice,mitrovicamilne,kendall}.
 Let $SL^{i}_{0}$ be the $i$th estimate of initial sea level and $SL^{i}(t_{p})$  the corresponding present-day
prediction. We then update the initial sea level by setting
\begin{equation}
  \label{eq:SL0it}
  SL_{0}^{i+1} = SL^{i}_{0} - [  SL^{i}(t_{p}) - SL_{p}].
\end{equation}

An alternative approach for determining the initial sea level has been recently
discussed by \cite{crawford2019viscoelastic} and \cite{lloyd2023}. Within this method, the objective functional
\begin{equation}
  \label{eq:Jinit}
  J = \frac{1}{2} \int_{\partial M} [SL(t_{p}) - SL_{p}]^{2} \dd S,
\end{equation}
is minimised with respect to $SL_{0}$ using gradient-based optimisation. Applying the results of this paper,
we can see that the initial sea level kernel for this objective functional takes the form
\begin{equation}
  D_{SL_{0}}\hat{J} = [SL(t_{p}) - SL_{p}] -
  \rho_{w} g\int_{\mathcal{I}} (C-C_{0}) (\mathcal{T}SL_{1}^{\dagger}) \dd t,
\end{equation}
where $SL_{1}^{\dagger}$ is obtained by solving the adjoint GIA problem  subject to 
\begin{equation}
  D_{SL}J =  [SL(t_{p}) - SL_{p}] \,\delta(t- t_{p}),
\end{equation}
defining  the only non-zero  force. If shoreline migration is neglected, then the second term
in $ D_{SL_{0}}\hat{J} $ vanishes, and this kernel is equal to the
negative of the update to the initial sea level within eq.(\ref{eq:SL0it}). It follows that
the traditional iterative approach for matching present-day sea level can be viewed as
an approximation to the steepest-descent minimisation of eq.(\ref{eq:Jinit}).

The constraint of matching the present-day sea level can be built directly into the definition of the
GIA forward problem. It is then a relatively
simple matter to reformulate the adjoint problem. We first write down a
modified Lagrangian
\begin{eqnarray}
  J + \int_{\mathcal{I}} \!\!\!\!\!\!&&\!\!\!\!\!\! \left\{
  \mathcal{A}(\mathbf{u},\phi \,| \, \mathcal{T} \mathbf{u}^{\dagger}, \mathcal{T}\phi^{\dagger})
  - \int_{M} 2 \mu \mathbf{m} : (\mathcal{T}\mathbf{d}^{\dagger}) \dd^{3} \mathbf{x}
  + \int_{\partial M} \left[(\mathcal{T}\mathbf{u}^{\dagger})\cdot \nabla \Phi +
    \mathcal{T}\phi^{\dagger} + \mathcal{T}\psi^{\dagger}\right] \,\sigma \dd S
  \right. \nonumber \\ &&+\int_{M}\rho \left[(\mathcal{T}\mathbf{u}^{\dagger})
    \cdot \nabla \psi + \mathbf{u} \cdot \nabla (\mathcal{T}\psi^{\dagger}) \right]
  \dd^{3}\mathbf{x} + (\mathcal{T}\bm{\omega}^{\dagger}) \cdot \mathbf{D}\cdot
  \bm{\omega} + \int_{M} 2\mu\,(\tau \,\dot{\mathbf{m}} +
  \mathbf{m}-\mathbf{d}):(\mathcal{T}\mathbf{m}^{\dagger}) \dd^{3}\mathbf{x}
  \nonumber \\ && + \int_{\partial M} \left[ \sigma - \rho_{w} \, C\, SL - \rho_{i}\,(1-C)\,I + \rho_{w} \,
  C_{0}\, SL_{0} + \rho_{i}\,(1-C_{0})\,I_{0}\right] g \,
  (\mathcal{T}SL_{1}^{\dagger}) \dd S \nonumber \\ && \left. + \int_{\partial M}\left( g\, SL_{1} + \mathbf{u}\cdot \nabla \Phi + \phi + \psi - \Phi_{g}
  \right) (\mathcal{T}\sigma^{\dagger}) \dd S - (\mathcal{T}\Phi_{g}^{\dagger})
  \int_{\partial M} \sigma \dd S \right\} \dd t \nonumber \\ && - \int_{\partial M}\rho_{w} g [SL(t_{p}) - SL_{p}] SL_{0}^{\dagger} \dd S.
\end{eqnarray}
The only differences from the Lagrangian in eq.(\ref{eq:Lag1}) are that $SL_{0}$ is here
 a state variable and the addition of the final term  -- with  a corresponding 
 Lagrange multiplier, $SL_{0}^{\dagger}$ --  associated with
the constraint  $SL(t_{p}) = SL_{p}$. 
Derivation of the adjoint equations  proceeds almost as before,
with  only the variation with respect to sea level being modified. To do this, it is useful
to decompose the sea level as $SL = SL_{0} + SL_{1}$ and  vary the two terms separately.
By letting $L$  be stationary with respect to $SL_{1}$ we obtain
\begin{equation}
  \sigma^{\dagger} = \rho_{w}  C^{\dagger} SL_{1}^{\dagger}
  -\frac{1}{g} \mathcal{T}D_{SL}J + \rho_{w} SL_{0}^{\dagger} \,\delta(t_{1} - t  + t_{0}-t_{p}),
\end{equation}
while by varying $SL_{0}$ we find
\begin{equation}
  SL_{0}^{\dagger} = \int_{\mathcal{I}}(C^{\dagger}-C_{0}) SL_{1}^{\dagger} \dd t
  - \frac{1}{g}\int_{I}D_{SL}J \dd t.
\end{equation}
Combining the two results,  we see that the adjoint load  takes the form
\begin{equation}
  \sigma^{\dagger} = \rho_{w}  C^{\dagger} SL_{1}^{\dagger}   -\frac{1}{g} \mathcal{T}D_{SL}J
  + \rho_{w}\left[\int_{\mathcal{I}} (C^{\dagger} - C_{0}) SL_{1}^{\dagger}  \dd t
    - \frac{1}{g}\int_{I}D_{SL}J \dd t\right]\delta(t_{1} - t+ t_{0}-t_{p}),
\end{equation}
while all other adjoint equations are unchanged.  The singular  part of the adjoint load
at the time-reversed present depends on the adjoint sea level
over the whole time-interval, and so these adjoint equations are not
amenable to a direct time-integration. However, a simple iterative scheme could
be applied, with an initial value for $SL_{0}^{\dagger}$ guessed and then
successively refined.   Expressions for the
sensitivity kernels for viscosity and ice thickness are unchanged, but now they
incorporate an implicit change in the initial sea level such that the
calculated present-day sea level always matches $SL_{p}$.

\section{Hessian kernels and the second-order adjoint GIA problem}

Within this section we apply the second-order adjoint method to the GIA
problem. These results extend earlier discussions within the literature that
have considered only the first-order theory. The purpose of the second-order
adjoint method is that it allows for second-derivatives of an objective
functional to be calculated at a practicable cost. In particular, the action of
the Hessian operator on a given model perturbation can be determined at a cost
equivalent to four GIA simulations. Such calculations are necessary within the
application of efficient Newton-type optimisation schemes, and also within a range of
methods for    uncertainty quantification for non-linear inverse
problems.

\subsection{Hessian operators and kernels}

Consider again an objective functional, $J$, and its reduced form, $\hat{J}$,
defined in terms of the model parameters in the GIA forward problem. To
simplify notations, we initially take viscosity, $\eta$, to be the only model
parameter. For a given viscosity perturbation, the second-order functional
derivative of $\hat{J}$ can be defined through the following Taylor expansion
\begin{equation}
  \hat{J}(\eta + \delta \eta) = \hat{J}(\eta) +
  \int_{M} D_{\eta}\hat{J}(\mathbf{x}) \,\delta \eta (\mathbf{x}) \dd^{3} \mathbf{x}
  + \frac{1}{2} \int_{M} \int_{M} D^{2}_{\eta\eta}\hat{J}(\mathbf{x},\mathbf{x}') \,\delta \eta (\mathbf{x}')
  \,\delta \eta(\mathbf{x}) \dd^{3} \mathbf{x}'  \dd^{3}\mathbf{x} + \cdots,
\end{equation}
where spatial arguments have been included for clarity.
Equivalently,  the second-order functional derivative can be defined through
the following limit
\begin{equation}
  \int_{M} \int_{M} D^{2}_{\eta\eta}\hat{J}(\mathbf{x},\mathbf{x}') \,\Delta \eta (\mathbf{x}')
  \,\delta \eta(\mathbf{x}) \dd^{3} \mathbf{x}'  \dd^{3}\mathbf{x}
  = \lim_{r\rightarrow 0} \lim_{s\rightarrow 0} \frac{1}{rs }\left[
    \hat{J}(\eta + r\, \delta \eta + s\, \Delta \eta )- \hat{J}(\eta)
    \right],
\end{equation}
where $\delta \eta$ and $\Delta \eta$ are arbitrary viscosity perturbations. This
latter definition makes it clear that $D^{2}_{\eta\eta}\hat{J}$ is symmetric
in its spatial arguments. If we define a new functional
\begin{equation}
  \hat{J}_{2}(\eta) =  \lim_{s\rightarrow 0} \frac{1}{s }\left[
    \hat{J}(\eta + s\, \Delta \eta  )- \hat{J}(\eta)
    \right] = \int_{M} D_{\eta}\hat{J}\, \Delta \eta \dd^{3}\mathbf{x},
\end{equation}
then we have
\begin{equation}
  \int_{M} D_{\eta}\hat{J}_{2} \,\delta \eta \dd^{3}\mathbf{x}' = \lim_{r\rightarrow 0} \frac{1}{r }\left[
    \hat{J}_{2}(\eta + r\, \delta \eta  )- \hat{J}_{2}(\eta)
    \right] = \int_{M} \int_{M} D^{2}_{\eta\eta}\hat{J}(\mathbf{x},\mathbf{x}') \,\Delta \eta (\mathbf{x}')
  \,\delta \eta(\mathbf{x}) \dd^{3} \mathbf{x}'  \dd^{3}\mathbf{x},
\end{equation}
which implies
\begin{equation}
  \label{eq:J2Id}
  D_{\eta}\hat{J}_{2}(\mathbf{x}) = \int_{M} D^{2}_{\eta\eta}\hat{J}(\mathbf{x},\mathbf{x}') \,\Delta \eta (\mathbf{x}')
  \dd^{3} \mathbf{x}'.
\end{equation}

The \emph{Hessian operator}, $\hat{\mathcal{H}}_{\eta\eta}$, of $\hat{J}$ at $\eta$ acts
on a viscosity perturbation, $\Delta \eta$, through
\begin{equation}
  (\hat{\mathcal{H}}_{\eta\eta} \,\Delta \eta)(\mathbf{x} ) = \int_{M} D^{2}_{\eta\eta}\hat{J}(\mathbf{x},\mathbf{x}') \,\Delta \eta (\mathbf{x}')
  \dd^{3} \mathbf{x}',
\end{equation}
with the result lying in the space of viscosity perturbations.
Due to the symmetry of $D^{2}_{\eta\eta}\hat{J}$,
we see that  $\hat{\mathcal{H}}_{\eta\eta}$ is self-adjoint.
Given this definition, the second-order Taylor expansion of
$\hat{J}$ takes the form
\begin{equation}
  \hat{J}(\eta + \delta \eta) = \hat{J}(\eta) +
  \int_{M} D_{\eta}\hat{J}(\mathbf{x}) \,\delta \eta (\mathbf{x}) \dd^{3} \mathbf{x}
  + \frac{1}{2} \int_{M} (\hat{\mathcal{H}}_{\eta\eta}\,\delta \eta)(\mathbf{x})
  \,\delta \eta(\mathbf{x})  \dd^{3}\mathbf{x} + \cdots,
\end{equation}
where the second-order term is expressed as  an integral of  $\delta \eta$
against the function   $\hat{\mathcal{H}}_{\eta\eta}\,\delta \eta$; we call this latter term the  \emph{Hessian kernel} of $\hat{J}$  relative to $\delta \eta$ \citep[c.f.][]{fichtner2011hessian}.
From eq.(\ref{eq:J2Id}), we see that the Hessian kernel relative to  $\Delta \eta$ is equal to the
functional derivative of $\hat{J}_{2}$, with this identity being key to our discussion of
second-order adjoint methods below.

These definitions extend readily to allow for the other model parameters within
the GIA forward problem. The full Hessian operator has a block structure
\begin{equation}
  \hat{\mathcal{H}} = \left(
  \begin{array}{ccc}
      \hat{\mathcal{H}}_{\eta\eta}   & \hat{\mathcal{H}}_{\eta I }   & \hat{\mathcal{H}}_{\eta SL_{0}}   \\
      \hat{\mathcal{H}}_{I \eta}     & \hat{\mathcal{H}}_{I  I }     & \hat{\mathcal{H}}_{I  SL_{0}}     \\
      \hat{\mathcal{H}}_{SL_{0}\eta} & \hat{\mathcal{H}}_{SL_{0} I } & \hat{\mathcal{H}}_{SL_{0} SL_{0}} \\
    \end{array}
  \right).
\end{equation}
Here, for example, $\hat{\mathcal{H}}_{I \eta}$ is a linear operator mapping viscosity perturbations
into ice thickness perturbations. Again, $\hat{\mathcal{H}}$ is self-adjoint,
this implying identities such as $\hat{\mathcal{H}}_{\eta I} = \hat{\mathcal{H}}_{ I \eta}^{*} $ with superscript,  $*$, denoting
the operator adjoint. The Hessian kernels (one for each model parameter) relative to
a given model perturbation, ($\Delta \eta$, $\Delta I$, $\Delta SL_{0}$), are defined in the obvious manner,
and can be identified with  the functional derivatives of
\begin{equation}
  \label{eq:J2Def}
  \hat{J}_{2} = \int_{M} D_{\eta}\hat{J} \, \Delta \eta \dd^{3} \mathbf{x}
  + \int_{\mathcal{I}} \int_{\partial M} D_{I}\hat{J} \, \Delta I \dd S \dd t + \int_{\partial M} D_{SL_{0}}\hat{J} \,\Delta SL_{0} \dd S.
\end{equation}

\subsection{Second-order PDE constrained optimisation through
  the Lagrangian method}

To derive the second-order adjoint equations, we follow the method of    \cite{syvret2022theoretical}.
With $L$ the Lagrangian in eq.(\ref{eq:Lag1}), and for fixed model
perturbations, $(\Delta \eta, \Delta I, \Delta SL_{0})$, we can define a new
functional
\begin{equation}
  J_{2} =
  \int_{M} D_{\eta}L \, \Delta \eta \dd^{3} \mathbf{x}
  + \int_{\mathcal{I}} \int_{\partial M} D_{I}L \, \Delta I \dd S \dd t + \int_{\partial M} D_{SL_{0}}L \,\Delta SL_{0} \dd S,
\end{equation}
 depending on the state variables ($\mathbf{u}$, $\phi$, $\bm{\omega}$, $\mathbf{m}$,
$SL$, $\sigma$, $\Phi_{g}$), the adjoint state variables   ($\mathbf{u}^{\dagger}$, $\phi^{\dagger}$,
$\bm{\omega}^{\dagger}$, $\mathbf{m}^{\dagger}$,
$SL^{\dagger}$, $\sigma^{\dagger}$, $\Phi_{g}^{\dagger}$), and the model parameters
($\eta$, $I$, $SL_{0}$). Using the results of the first-order adjoint theory,
this functional can be written explicitly as
\begin{eqnarray}
  J_{2} &\!\!\!\!=\!\!\!\!&
  \int_{\mathcal{I}} \int_{M}\frac{2\mu }{\eta}(\mathbf{m} - \mathbf{d}):
  (\mathcal{T}\mathbf{m}^{\dagger}) \,\Delta \eta \dd^{3}\mathbf{x} \dd t -
  \int_{\mathcal{I}} \int_{\partial M} \rho_{i} g\left[ (1-C) \,\Delta I - (1-C_{0})\,\Delta I_{0}
  \right](\mathcal{T} SL_{1}^{\dagger}) \dd S \dd t \nonumber \\ && +
  \int_{\mathcal{I}} \int_{\partial M} \left[ D_{SL}J - \rho_{w}g\,(C-C_{0})
  (\mathcal{T}SL_{1}^{\dagger})\right] \Delta SL_{0} \dd S \dd t.
\end{eqnarray}
By requiring that  the arguments of $J_{2}$ satisfy both the forward and first-order
adjoint GIA problems, we have the equality $J_{2} = \hat{J}_{2}$ with $\hat{J}_{2}$ defined
in eq.(\ref{eq:J2Def}).
It follows that the
Hessian kernels of $\hat{J}$  relative to the chosen model perturbations
can be obtained by differentiating $J_{2}$  subject to these constraints. To proceed, we define
a new Lagrangian
\begin{eqnarray}
  L_{2} = J_{2} +
  \int_{\mathcal{I}} \!\!\!\!\!\!&&\!\!\!\!\!\! \left\{
  \mathcal{A}(\mathbf{u},\phi \,| \, \mathcal{T} \Delta \mathbf{u}^{\dagger},
  \mathcal{T}\Delta \phi^{\dagger}) - \int_{M} 2 \mu \mathbf{m} :
  (\mathcal{T}\Delta \mathbf{d}^{\dagger}) \dd^{3} \mathbf{x} + \int_{\partial M} \left[(\mathcal{T}\Delta \mathbf{u}^{\dagger})\cdot \nabla \Phi +
    \mathcal{T}\Delta \phi^{\dagger} + \mathcal{T}\Delta\psi^{\dagger}\right]
  \,\sigma \dd S \right. \nonumber \\ &&+\int_{M}\rho \left[(\mathcal{T}\Delta
    \mathbf{u}^{\dagger}) \cdot \nabla \psi + \mathbf{u} \cdot \nabla
    (\mathcal{T}\Delta \psi^{\dagger}) \right] \dd^{3}\mathbf{x} +
  (\mathcal{T}\Delta \bm{\omega}^{\dagger}) \cdot \mathbf{D}\cdot \bm{\omega} +
  \int_{M} 2\mu\,(\tau \,\dot{\mathbf{m}} +
  \mathbf{m}-\mathbf{d}):(\mathcal{T}\Delta \mathbf{m}^{\dagger})
  \dd^{3}\mathbf{x} \nonumber \\ && + \int_{\partial M} \left[ \sigma - \rho_{w} \, C\, SL - \rho_{i}\,(1-C)\,I + \rho_{w} \,
  C_{0}\, SL_{0} + \rho_{i}\,(1-C_{0})\,I_{0}\right] g \, (\mathcal{T}\Delta
  SL_{1}^{\dagger}) \dd S \nonumber \\ && \left. + \int_{\partial M}\left( g\, SL_{1} + \mathbf{u}\cdot \nabla \Phi + \phi + \psi - \Phi_{g}
  \right) (\mathcal{T}\Delta \sigma^{\dagger}) \dd S - (\mathcal{T}\Delta
  \Phi_{g}^{\dagger}) \int_{\partial M} \sigma \dd S \right\} \dd t \nonumber \\ &&+
  \mathcal{A}(\mathbf{u}^{\dagger},\phi^{\dagger}\,|\, \mathcal{T}\Delta
  \mathbf{u}, \mathcal{T}\Delta \phi) - \int_{M} 2\mu \, \mathbf{m}^{\dagger}:
  (\mathcal{T}\Delta \mathbf{d}) \dd^{3}\mathbf{x} + \int_{\partial M}\left[ (\mathcal{T}\Delta \mathbf{u})\cdot \nabla \Phi +
    \mathcal{T}\Delta \phi + \mathcal{T}\Delta \psi \right] \sigma^{\dagger} \dd S
  \nonumber \\ && +\int_{\partial M} \left[\left(\mathcal{T} D_{\mathbf{u}}J \right)\cdot
    (\mathcal{T}\Delta \mathbf{u}) + \left(\mathcal{T} D_{\phi}J\right)\,(\mathcal{T}\Delta \phi)\right] \dd S + \int_{M}\rho\,\left[
    \mathbf{u}^{\dagger}\cdot \nabla (\mathcal{T}\Delta \psi) + (\mathcal{T}\Delta
    \mathbf{u})\cdot \nabla \psi^{\dagger}\right] \dd^{3}\mathbf{x} \nonumber \\ &&
  + (\mathcal{T}\Delta \bm{\omega}) \cdot \left[\mathbf{D}\cdot
    \bm{\omega}^{\dagger} + \left(\mathcal{T}D_{\bm{\omega}}J \right)\right] + \int_{M}2\mu \left( \tau
  \dot{\mathbf{m}}^{\dagger} + \mathbf{m}^{\dagger} -
  \mathbf{d}^{\dagger}\right):(\mathcal{T}\Delta \mathbf{m}) \dd^{3}\mathbf{x}
  \nonumber \\ && + \int_{\partial M} \left[ \sigma^{\dagger} -\rho_{w}\, C^{\dagger}\, SL^{\dagger}_{1} +
    \frac{1}{g} \mathcal{T}D_{SL}J \right](g \mathcal{T}\Delta
  SL_{1}) \dd S \nonumber \\ && \left. + \int_{\partial M} \left( g\, SL_{1}^{\dagger} + \mathbf{u}^{\dagger}\cdot \nabla \Phi +
  \phi^{\dagger} + \psi^{\dagger} - \Phi_{g}^{\dagger} \right) \,(\mathcal{T}
  \Delta \sigma) \dd S - (\mathcal{T}\Delta \Phi_{g})\int_{\partial M} \sigma^{\dagger} \dd S \right\} \dd t.
\end{eqnarray}
Here, the \emph{second-order adjoint state variables} ($\Delta \mathbf{u}^{\dagger}$, $\Delta\phi^{\dagger}$, $\Delta\bm{\omega}^{\dagger}$, $\Delta\mathbf{m}^{\dagger}$,
$\Delta SL_{1}^{\dagger}$, $\Delta\sigma^{\dagger}$, $\Delta\Phi_{g}^{\dagger}$) are associated with
the constraint that the state variables solve the forward GIA problem, while the  \emph{second-order state
  variables}  ($\Delta \mathbf{u}$, $\Delta\phi$, $\Delta\bm{\omega}$, $\Delta\mathbf{m}$,
$\Delta SL_{1}$, $\Delta\sigma$, $\Delta\Phi_{g}$) do the same but for the first-order adjoint
problem. Note that both sets of new variables have been time-reversed for convenience. The
Lagrange multiplier theorem tells us that the functional derivatives of $L_{2}$ and $\hat{J}_{2}$
with respect to the model parameters coincide so long as $L_{2}$ is stationary with respect to
its other variables. Varying the second-order state and adjoint state variables
gives the forward and first-order adjoint problems. Requiring that
$L_{2}$ is stationary with respect to the state and first-order adjoint state variables
then gives two new sets of equations that the second-order state and adjoint state  variables must satisfy.

By varying $L_{2}$  with respect to the first-order adjoint state variables we obtain 
\begin{eqnarray}
  && \mathcal{A}(\Delta \mathbf{u}, \Delta \phi \,|\, \delta \mathbf{m}^{\dagger}, \delta \phi^{\dagger})
  - \int_{M} 2\mu \Delta \mathbf{m} : \delta \mathbf{d}^{\dagger} \dd^{3}\mathbf{x}
  + \int_{\partial M} \left(\delta \mathbf{u}^{\dagger} \cdot \nabla \Phi + \delta
  \phi^{\dagger} + \delta \psi^{\dagger} \right) \Delta \sigma \dd S \nonumber \\
  && + \int_{M} \rho \left( \delta \mathbf{u}^{\dagger} \cdot \nabla \Delta \psi
  + \Delta \mathbf{u} \cdot \nabla \delta \psi^{\dagger} \right) \dd^{3}
  \mathbf{x} + \delta \bm{\omega}^{\dagger} \cdot \mathbf{D} \cdot \Delta
  \bm{\omega} + \int_{M} 2\mu \left[ \tau \Delta \dot{\mathbf{m}} + \Delta
    \mathbf{m} - \Delta \mathbf{d} + \frac{\Delta
      \eta}{\eta}(\mathbf{m}-\mathbf{d}) \right] :\delta \mathbf{m}^{\dagger}
  \dd^{3}\mathbf{x} = 0, \\ && \Delta \sigma = \rho_{w} C \Delta SL_{1} +
  \rho_{w} (C-C_{0}) \,\Delta SL_{0} + \rho_{i}\left[(1-C)\,\Delta I -
  (1-C_{0})\,\Delta I_{0}\right], \\ && \Delta SL_{1} = -\frac{1}{g}\left( \Delta
  \mathbf{u} \cdot \nabla \Phi + \Delta \phi + \Delta \psi \right) + \frac{\Delta
    \Phi_{g}}{g}, \\ && \int_{\partial M} \Delta \sigma \dd S = 0.
\end{eqnarray}
where in the first equation ($\delta \mathbf{u}^{\dagger}$, $\delta \phi^{\dagger}$, $\delta \bm{\omega}^{\dagger}$,
$\delta \mathbf{m}^{\dagger}$) are arbitrary time-independent test functions,
along with the initial condition, $\Delta \mathbf{m}(\mathbf{x},t_{0}) = \mathbf{0}$.
These equations can be seen to be the linearisation
of the forward GIA problem with respect to the model parameters, this being
a general feature of the second-order adjoint method. We note that the
linearisation of the GIA forward problem with respect to
its model parameters has also been discussed by
\cite{martinec2015forward} in the context of what they term
the ``forward sensitivity method''.

Calculating the variation of $L_{2}$ with respect to the state variables is
more involved due the dependence of the forces within the first-order adjoint
equations on the state variables. For example, consider the term
\begin{equation}
  \int_{\mathcal{I}} \int_{\partial M} \left(\mathcal{T} D_{\mathbf{u}}J \right)\cdot
  (\mathcal{T}\Delta \mathbf{u}) \dd^{3}\mathbf{x} \dd t = \int_{\mathcal{I}} \int_{\partial M} D_{\mathbf{u}}J \cdot \Delta \mathbf{u}
  \dd^{3}\mathbf{x} \dd t,
\end{equation}
with the  equality following from the properties of $\mathcal{T}$.
If we vary this functional with respect to $\mathbf{u}$, we obtain
\begin{equation}
  \int_{\mathcal{I}} \int_{\partial M} (\mathcal{H}_{\mathbf{u}\mathbf{u}} \Delta \mathbf{u}) \cdot \delta \mathbf{u}
  \dd^{3}\mathbf{x} \dd t,
\end{equation}
where we have recalled the definition of the Hessian operator of a functional; note that the Hessian of
$\hat{J}$ has a hat on, but that for $J$ does not. This idea
extends readily to the variations of this functional with respect to other state variables, and to the
other adjoint forces within $L_{2}$. Using these notations, the variation
of $L_{2}$ with respect to the state variables yields the following set of equations
\begin{eqnarray}
  && \mathcal{A}(\Delta \mathbf{u}^{\dagger}, \Delta \phi^{\dagger}
  \,|\, \delta \mathbf{m},
  \delta \phi)
  - \int_{M} 2\mu \left(\Delta \mathbf{m}^{\dagger}-\frac{\Delta \eta}{\eta} \mathbf{m}^{\dagger}\right)
  : \delta \mathbf{d} \dd^{3}\mathbf{x}
  + \int_{\partial M} \left(\delta \mathbf{u} \cdot \nabla \Phi + \delta \phi + \delta \psi
  \right) \Delta \sigma^{\dagger} \dd S \nonumber \\ && + \int_{M} \rho \left(
  \delta \mathbf{u} \cdot \nabla \Delta \psi^{\dagger} + \Delta
  \mathbf{u}^{\dagger} \cdot \nabla \delta \psi \right) \dd^{3} \mathbf{x} +
  \delta \bm{\omega} \cdot \mathbf{D} \cdot \Delta \bm{\omega}^{\dagger} +
  \int_{M} 2\mu \left[ \tau \Delta \dot{ \mathbf{m}}^{\dagger} + \Delta
    \mathbf{m}^{\dagger} - \Delta \mathbf{d}^{\dagger} + \frac{\Delta \eta}{\eta}
    \mathbf{m}^{\dagger} \right] :\delta \mathbf{m} \dd^{3}\mathbf{x} \nonumber \\
  && + \int_{\partial M} \left( \mathcal{T} \mathcal{H}_{\mathbf{u}\mathbf{u}} \,\Delta \mathbf{u} +
  \mathcal{T} \mathcal{H}_{\mathbf{u}\phi} \,\Delta \phi + \mathcal{T}
  \mathcal{H}_{\mathbf{u}\bm{\omega}} \,\Delta \bm{\omega} \right) \cdot \delta \mathbf{u}
  \dd S + \int_{\partial M} \left( \mathcal{T} \mathcal{H}_{\phi\mathbf{u}} \,\Delta \mathbf{u} + \mathcal{T}
  \mathcal{H}_{\phi\phi} \,\Delta \phi + \mathcal{T} \mathcal{H}_{\phi\bm{\omega}} \,\Delta
  \bm{\omega} \right) \delta \phi \dd S \nonumber \\ && + \left( \mathcal{T}
  \mathcal{H}_{\bm{\omega}\mathbf{u}} \,\Delta \mathbf{u} + \mathcal{T} \mathcal{H}_{\bm{\omega}\phi}
  \,\Delta \phi + \mathcal{T} \mathcal{H}_{\bm{\omega}\bm{\omega}} \,\Delta \bm{\omega}
  \right) \cdot \delta \bm{\omega} = 0, \\ && \Delta \sigma^{\dagger} = \rho_{w}
  C^{\dagger} \,\Delta SL_{1}^{\dagger} - \rho_{w} \mathcal{T}\!\left(
  \frac{\rho_{w} \Delta SL - \rho_{i} \Delta I}{\rho_{w}
    \partial_{\perp}SL - \rho_{i}\partial_{\perp}I} \delta_{\mathcal{O}_{t}} \!\right) SL_{1}^{\dagger} \nonumber \\
  && \qquad\qquad - \frac{1}{g} \mathcal{T} \left(\mathcal{H}_{SL SL} \Delta SL + \mathcal{H}_{SL
    \mathbf{u}} \Delta \mathbf{u} + \mathcal{H}_{SL \phi} \Delta \phi + \mathcal{H}_{SL \bm{\omega}}
  \Delta \bm{\omega}\right), \\ && \Delta SL_{1}^{\dagger} = -\frac{1}{g}\left(
  \Delta \mathbf{u}^{\dagger} \cdot \nabla \Phi + \Delta \phi^{\dagger} + \Delta
  \psi^{\dagger} \right) + \frac{\Delta \Phi_{g}^{\dagger}}{g}, \\ && \int_{\partial M} \Delta \sigma^{\dagger} \dd S = 0,
\end{eqnarray}
where ($\delta \mathbf{d}$, $\delta \phi$, $\delta \bm{\omega}$, $\delta m$)
are arbitrary time-independent test functions,
we have defined $\Delta SL = \Delta SL_{0} + \Delta SL_{1}$, and we have
the initial condition, $\Delta \mathbf{m}^{\dagger}(\mathbf{x},t_{0}) = \mathbf{0}$.

\subsection{Practical implementation}

Within the second-order adjoint theory there are four equations to be solved.
First, there is the forward GIA problem. Next, we have the first-order
adjoint equations which are linear but with coefficients and force terms
dependent on the solution of the forward problem. Solution of these first two
equations is, of course, necessary for the calculation of sensitivity kernels
via the first-order adjoint method. The final two equations determine the
second-order state and adjoint state variables, with both sets of equations
depending explicitly on the assumed model perturbations ($\Delta \eta$, $\Delta
  I$, $\Delta SL_{0}$). The equations for the second-order state variables
depend on the solution of the forward problem due to the occurrence of the
ocean function, $C$, but are independent of either set of adjoint
variables. Finally, the second-order adjoint variables can be determined, with
the force terms in this case depending on the solution of the state variables,
first-order adjoint state variables, and second-order state variables. The
sequence of dependencies just described means that the four equations can be
solved sequentially.

As with the first-order adjoint theory,
the new  equations to be solved resemble very closely the forward GIA problem,
with the principal difference being the occurrence of additional force terms.
Within the second-order equations  the additional forces within the elastostatic sub-problems
 include  volumetric components associated with the given viscosity perturbation, $\Delta \eta$.
Nevertheless, the modifications necessary to implement the second-order adjoint
theory within existing GIA codes are modest.

The uniqueness and existence of solutions to the equations for the second-order
state and adjoint state variables can be readily assessed. In particular,
within the equations for the second-order adjoint state variables, the
requirement that the additional force terms apply no net force nor torque to
the earth model  follows from the expansion of eq.(\ref{eq:Jinv}) to
second-order. Similarly, the second-order variables are defined only up to
rigid body motions, but it will be seen shortly that such terms have no effect
on the Hessian kernels.

Finally, we note that the equations for the
second-order adjoint state variables contain force terms that can be singular
in both space and time. As with the first-order adjoint equations, practical
methods for dealing with these singularities can be developed but will be
discussed  in a future work.

\subsection{Expressions for the Hessian kernels}

By differentiating the Lagrangian, $L_{2}$, with respect to the model
parameters, we can write the functional derivatives of $\hat{J}_{2}$ in terms
of the solution of the state and adjoint state variables. As noted previously,
these functional derivatives are equal to the components of the action of the
Hessian of $\hat{J}$ on the selected model perturbations. In the case of
viscosity the calculations are easy, with the result
\begin{eqnarray}
  D_{\eta}\hat{J}_{2} &\!\!\!\!=\!\!\!\!&
  \int_{\mathcal{I}}  \frac{ 2 \mu}{\eta } \left[
    (\mathbf{m}-\mathbf{d}) :  \mathcal{T} \left( \Delta \mathbf{m}^{\dagger}-\frac{  \Delta \eta }{\eta}  \mathbf{m}^{\dagger}\right)
    + (\mathbf{m}^{\dagger}- \mathbf{d}^{\dagger}) : (\mathcal{T} \Delta \mathbf{m}) \right] \dd t.
\end{eqnarray}
Functional derivatives with respect to
ice thickness and initial sea level are more complicated due to the
dependence of the ocean function on these fields, but
following a now routine calculation using eq.(A30) of \cite{crawford2018}
we find
\begin{eqnarray}
  D_{I}\hat{J}_{2} &\!\!\!\!=\!\!\!\!& - \rho_{i} g \left[
  (1-C) (\mathcal{T} \Delta SL_{1}^{\dagger})
  - (1-C_{0}) \int_{\mathcal{I}} \mathcal{T} \Delta SL_{1}^{\dagger} \dd t\,
  \delta(t-t_{0})
  \right] + \rho_{i} g  \frac{\rho_{w} \Delta SL - \rho_{i}
    \Delta I}{\rho_{w}\partial_{\perp}SL - \rho_{i}\partial_{\perp} I} (\mathcal{T} SL_{1}^{\dagger}) \, \delta_{\partial \mathcal{O}_{t}} \nonumber \\ && - \rho_{i} g \frac{\rho_{w} \Delta SL_{0}
  - \rho_{i} \Delta I_{0}}{\rho_{w}\partial_{\perp}SL_{0} - \rho_{i}\partial_{\perp} I_{0}} \int_{\mathcal{I}} \mathcal{T} SL_{1}^{\dagger} \dd t \,
  \delta_{\partial \mathcal{O}_{t_{0}}} \, \delta (t-t_{0}), \\ D_{SL_{0}}\hat{J}_{2} &\!\!\!\!=\!\!\!\!& \int_{\mathcal{I}} \left[
  \mathcal{H}_{SLSL} \Delta SL_{0} - \rho_{w} g (C-C_{0})(\mathcal{T} \Delta
  SL_{1}^{\dagger}) \right] \dd t - \int_{\mathcal{I}} \rho_{w} g \frac{\rho_{w}
    \Delta SL - \rho_{i} \Delta I}{\rho_{w}
    \partial_{\perp}SL - \rho_{i}\partial_{\perp} I} (\mathcal{T}SL_{1}^{\dagger}) \, \delta_{\mathcal{O}_{t}} \dd t
  \nonumber \\ && + \rho_{w} g \frac{\rho_{w} \Delta SL_{0} - \rho_{i} \Delta
  I_{0}}{\rho_{w}
  \partial_{\perp}SL_{0} - \rho_{i}\partial_{\perp} I_{0}} \int_{\mathcal{I}} \mathcal{T} SL_{1}^{\dagger} \dd t \,
  \delta_{\mathcal{O}_{t_{0}}},
\end{eqnarray}
where we recall that $\Delta SL = \Delta SL_{0} + \Delta SL_{1}$. By inspection, non-uniqueness
within the solution of the various elastostatic sub-problems does not 
propagate through to the Hessian kernels.

Looking at the above expressions, we note that certain terms within the Hessian kernels are independent of the
second-order variables. This leads to the following approximate formulae:
\begin{eqnarray}
  D_{\eta}\hat{J}_{2} &\!\!\!\!\approx\!\!\!\!&
 - \int_{\mathcal{I}}  \frac{ 2 \mu \Delta \eta}{\eta^{2} } 
    (\mathbf{m}-\mathbf{d}) :  (\mathcal{T}    \mathbf{m}^{\dagger})
 \dd t, \\
   D_{I}\hat{J}_{2} &\!\!\!\!\approx\!\!\!\!&  - \rho_{i} g \frac{\rho_{w} \Delta SL_{0}
  - \rho_{i} \Delta I_{0}}{\rho_{w}\partial_{\perp}SL_{0} - \rho_{i}\partial_{\perp} I_{0}} \int_{\mathcal{I}} \mathcal{T} SL_{1}^{\dagger} \dd t \,
  \delta_{\partial \mathcal{O}_{t_{0}}} \, \delta (t-t_{0}), \\ D_{SL_{0}}\hat{J}_{2} &\!\!\!\!\approx\!\!\!\!& \int_{\mathcal{I}} 
  \mathcal{H}_{SLSL} \Delta SL_{0} \dd t + \rho_{w} g \frac{\rho_{w} \Delta SL_{0} - \rho_{i} \Delta
  I_{0}}{\rho_{w}
  \partial_{\perp}SL_{0} - \rho_{i}\partial_{\perp} I_{0}} \int_{\mathcal{I}} \mathcal{T} SL_{1}^{\dagger} \dd t \,
  \delta_{\mathcal{O}_{t_{0}}},
\end{eqnarray}
that could be used to approximate the action of the Hessian within the context of a quasi-Newton optimisation scheme 
without having to solve for the second-order variables.

\section{Discussion}

The aims of this paper have been three-fold. First, we have extended adjoint GIA theory to account for rotational feedbacks. This aspect of the problem was (knowingly) neglected within earlier discussions, but it is sufficiently important that it must be included in any serious practical application. The second was to present a version
of the adjoint GIA equations that is independent of the rate-formulation used by \cite{alattartromp} and \cite{ crawford2018}. While the rate-formulation has its merits,
its unfamiliarity has probably limited the adoption of adjoint methods within
the GIA community. Moreover, the apparent need for explicit time-stepping schemes severely limits its numerical efficiency. Finally, details of the second-order
adjoint theory for the GIA problem have been documented in full. Within future work on the GIA inverse problem, a key issue to address is uncertainty within the generated models. Second-order
adjoint methods underlie  the  few techniques available   for uncertainty quantification within large-scale non-linear inverse problems, and hence we expect that
there will be growing need for these results in the years to come.

\begin{acknowledgments}
  We thank Jerry Mitrovica, Jacky Austermann, and Will Eaton for helpful comments and suggestions.
  Natural Environment Research Council grant numbers NE/V010433/1  provided support for DA \& FS. 
  Natural Environment Research Council grant numbers NE/X013804/1  provided support for DA \& ZY.
  National Science Foundation  grants NSF-EAR-2002352 and OPP-2142592 provided support for AL.
  
\end{acknowledgments}

\section*{Data availability statement}

There is no data associated with this paper.

\bibliographystyle{gji}
\bibliography{references}

\appendix

\section{Accounting for a fluid core}

\subsection{Geometry of the earth model}

As in the main text, we let $M$ denote the volume occupied by the equilibrium earth model. We now suppose that this set is decomposed as $M_{S} \cup M_{F}$ where
$M_{S}$ denotes solid regions and $M_{F}$  fluid. The precise number of solid and fluid regions need not be specified, but we assume that
the different regions are nested one within another, and that the outermost region is solid. We write $\Sigma$ for the union of all internal and
external boundaries, with this set decomposed as $\Sigma = \partial M \cup \Sigma_{FS} \cup \Sigma_{SF}$, where $\Sigma_{FS}$ (resp. $\Sigma_{SF}$) denotes
boundaries between fluid and solid regions where the fluid is on the inner (resp. outer) side of the boundary. Note that internal boundaries between solid regions are permitted, representing discontinuities
in physical parameters, but such boundaries require no special consideration within the equations of motion and so will not be explicitly discussed. 

\subsection{Elastostatic problem including fluid regions}

To model elastostatic deformation in such an earth model  we follow \cite{dahlen74} who showed
 that the linearised Lagrangian  displacement cannot be well-defined within fluid regions, but
that an Eulerian formulation can instead be used. Moreover, due to the hydrostatic equilibrium condition 
within fluid regions, all necessary dynamical fields can be expressed in terms of the gravitational potential perturbation.
 \cite{bagheri2019tidal} generalised Dahlen's arguments  slightly to account for
an applied tidal potential, showing that perturbations to density and pressure within fluid regions are given by
\begin{equation}
  \rho_{1} = g^{-1} \partial_{r}\rho \,(\phi+\psi), \quad p_{1} = -\rho\,(\phi+\psi). 
\end{equation}
Using these results, eq.(\ref{eq:EQM2}) is generalised to
\begin{eqnarray}
  \label{eq:EQM2F}
&&  \mathcal{A}(\mathbf{u},\phi \,| \, \mathbf{u}', \phi') + \int_{\partial M} (\mathbf{u}'\cdot \nabla \Phi + \phi') \,\sigma \dd S +\int_{M_{S}}\rho \,
  \mathbf{u}' \cdot \nabla \psi \dd^{3}\mathbf{x}  \nonumber \\ && + \int_{M_{F}} g^{-1}\partial_{r}\rho\, \psi \,\phi' \dd^{3}\mathbf{x}
  +\int_{\Sigma_{FS}}\rho^{-} \psi \,\unvec\cdot\mathbf{u}' \dd S    -\int_{\Sigma_{FS}}\rho^{+} \psi \,\unvec\cdot\mathbf{u}' \dd S = 0, 
\end{eqnarray}
where $\rho^{\pm}$ denotes the density evaluated on the upper (+) or lower (-) side of a boundary. Here the bilinear form
$\mathcal{A}$ is now given by
\begin{eqnarray}
  \mathcal{A}(\mathbf{u},\phi \,| \, \mathbf{u}', \phi') &=& \int_{M_{S}} \kappa\, \nabla \cdot \mathbf{u} \, \nabla \cdot \mathbf{u}' \dd^{3}\mathbf{x}
  + \int_{M_{S}}2 \mu\, \mathbf{d}:\mathbf{d}' \dd^{3} \mathbf{x} + \frac{1}{2}\int_{M_{S}}\rho \left[
    \nabla(\mathbf{u}\cdot \nabla \Phi)\cdot \mathbf{u}' + \nabla(\mathbf{u}'\cdot \nabla \Phi)\cdot \mathbf{u}
    \right] \dd^{3} \mathbf{x} \nonumber \\
  && -\frac{1}{2}\int_{M_{S}}\rho \left(
  \nabla \cdot \mathbf{u} \, \nabla \Phi \cdot \mathbf{u}' + \nabla \cdot \mathbf{u}' \, \nabla \Phi \cdot \mathbf{u}
  \right) \dd^{3}\mathbf{x} + \int_{M_{S}} \rho \left(
  \nabla \phi  \cdot \mathbf{u}' + \nabla \phi'  \cdot \mathbf{u}
  \right) \dd^{3} \mathbf{x} \nonumber \\ && + \frac{1}{4\pi G}\int_{\mathbb{R}^{3}} \nabla \phi \cdot \nabla \phi' \dd^{3}\mathbf{x}
  + \int_{M_{F}}g^{-1} \partial_{r}\rho\,\phi \phi' \dd^{3}\mathbf{x} + \int_{\Sigma_{FS}} \rho^{-} g\, \unvec\cdot \mathbf{u} \, \unvec\cdot \mathbf{u}' \dd S
  \nonumber \\ && - \int_{\Sigma_{SF}} \rho^{+} g\, \unvec\cdot \mathbf{u} \, \unvec\cdot \mathbf{u}' \dd S
  + \int_{\Sigma_{FS}}\rho^{-}(\phi \mathbf{u}' +\phi' \mathbf{u})\cdot \unvec \dd S - \int_{\Sigma_{SF}}\rho^{+}(\phi \mathbf{u}' +\phi' \mathbf{u})\cdot \unvec \dd S,
\end{eqnarray}
as derived within \cite{alattartromp}. Notably, the symmetry of the bilinear form in eq.(\ref{eq:Aadjoint}) is retained
along with eq.(\ref{eq:Akernel}) for linearised rigid body motions. It follows that our discussion of 
uniqueness and existence within the main paper remains valid.

\subsection{Rotational feedbacks}

Our earlier discussion of rotational feedbacks followed \cite{al2024reciprocity}, with the key identity being  
\begin{equation}
  \bm{\omega}' \cdot \mathbf{D}
\cdot \bm{\omega} + \int_{M} \rho\,\mathbf{u} \cdot \nabla\psi' \dd^{3}\mathbf{x} + \int_{\partial M} \sigma \,\psi' \dd S = 0,
\end{equation}
which serves to define $\bm{\omega}$ in terms of the load and the associated displacement vector. Derivation of this result
depends on the linearised relation between the displacement vector and the inertia tensor perturbation. Such a relation carries
over to solid regions of the model, but within fluid regions a modified formula is required. To do this, we write the
inertia tensor perturbation in fluid regions using Eulerian variables, with a volumetric contribution involving $\rho_{1}$ defined
above, and boundary perturbations expressed in terms of the displacement vector on the solid-side. The result is the generalised relation
\begin{eqnarray}
  \bm{\omega}' \cdot \mathbf{D} \cdot \bm{\omega} + \int_{M_{S}} \rho \, \mathbf{u} \cdot \nabla \psi' \dd^{3}\mathbf{x}
  + \int_{M_{F}} g^{-1} \partial_{r}\rho (\phi + \psi) \psi' \dd^{3}\mathbf{x} + \int_{\Sigma_{FS}} \rho \psi' \mathbf{u}\cdot \unvec
  \dd S - \int_{\Sigma_{SF}} \rho \psi' \mathbf{u}\cdot \unvec \dd S + \int_{\partial M} \sigma \,\psi' \dd S= 0, 
\end{eqnarray}
which is to hold for all $\bm{\omega}'$. This can be combined with the weak-form of the elastostatic problem to
generalise eq.(\ref{eq:EQM3}) to the case of fluid-solid earth models:
\begin{eqnarray}
    \label{eq:EQM3F}
&&  \mathcal{A}(\mathbf{u},\phi \,| \, \mathbf{u}', \phi') + \int_{\partial M} (\mathbf{u}'\cdot \nabla \Phi + \phi' + \psi') \,\sigma \dd S +\int_{M_{S}}\rho \,
    \left(\mathbf{u}' \cdot \nabla \psi + \mathbf{u} \cdot \nabla \psi' \right) \dd^{3}\mathbf{x}  
    + \int_{M_{F}} g^{-1}\partial_{r}\rho\, \left(\psi \,\phi' + \psi'\,\phi + \psi\,\psi'\right) \dd^{3}\mathbf{x} \nonumber \\ &&
  +\int_{\Sigma_{FS}}\rho^{-} \left(\psi \,\unvec\cdot\mathbf{u}' + \psi' \,\unvec\cdot\mathbf{u} \right)\dd S    -\int_{\Sigma_{FS}}\rho^{+} \left(\psi \,\unvec\cdot\mathbf{u}' + \psi' \,\unvec\cdot\mathbf{u} \right) \dd S
  +\bm{\omega}' \cdot \mathbf{D} \cdot \bm{\omega} = 0. 
\end{eqnarray}
The incorporation of viscoelasticity and sea level into the problem follows exactly as in the main text.

It would be easy, if tedious, to write down modified Lagrangians and carry through the full derivation of the first- and second-order
adjoint problems, but this is not necessary. Instead, we need merely note that both eq.(\ref{eq:EQM3}) and (\ref{eq:EQM3F}) involve
symmetric bilinear forms in the triplets $(\mathbf{u}, \phi,\bm{\omega})$ and $(\mathbf{u}', \phi',\bm{\omega}')$. It is this symmetry alone that
accounts for the elastostatic and rotational equations remaining unchanged between the forward and adjoint problems. Moreover, the model parameters of interest 
do not occur within the new terms linked to fluid regions, and hence  expressions for the sensitivity and Hessian kernels
in the main paper remain valid.

\end{document}